\documentclass[journal]{IEEEtran}
\IEEEoverridecommandlockouts
\usepackage{cite}
\usepackage{empheq}
\usepackage{amsthm}
\usepackage{amsmath,amssymb,amsfonts,eqnarray}
\usepackage{graphicx}
\usepackage{textcomp}
\usepackage{xcolor}
\usepackage{acronym}
\usepackage{tabularx}
\usepackage{algorithm}
\usepackage{algpseudocode}
\usepackage{multirow}
\usepackage{academicons}
\usepackage{scalerel}
\usepackage{tikz}
\usetikzlibrary{svg.path}
\usepackage{subcaption}
\usepackage{url}
\usepackage{ragged2e}
\usepackage{steinmetz}
\usepackage[symbol]{footmisc}
\usepackage{mathtools}
\usepackage{bm}
\usepackage{makecell}
\setcellgapes{4pt}
\usepackage{enumitem}
\usepackage{siunitx}

\newcommand{\norm}[1]{\left\lVert#1\right\rVert}
\def\BibTeX{{\rm B\kern-.05em{\sc i\kern-.025em b}\kern-.08em
    T\kern-.1667em\lower.7ex\hbox{E}\kern-.125emX}}

  \tikzset{
  orcidlogo/.pic={
    \fill[orcidlogocol] svg{M256,128c0,70.7-57.3,128-128,128C57.3,256,0,198.7,0,128C0,57.3,57.3,0,128,0C198.7,0,256,57.3,256,128z};
    \fill[white] svg{M86.3,186.2H70.9V79.1h15.4v48.4V186.2z}
                 svg{M108.9,79.1h41.6c39.6,0,57,28.3,57,53.6c0,27.5-21.5,53.6-56.8,53.6h-41.8V79.1z M124.3,172.4h24.5c34.9,0,42.9-26.5,42.9-39.7c0-21.5-13.7-39.7-43.7-39.7h-23.7V172.4z}
                 svg{M88.7,56.8c0,5.5-4.5,10.1-10.1,10.1c-5.6,0-10.1-4.6-10.1-10.1c0-5.6,4.5-10.1,10.1-10.1C84.2,46.7,88.7,51.3,88.7,56.8z};
  }
}

\newcommand\orcidicon[1]{\href{https://orcid.org/#1}{\mbox{\scalerel*{
\begin{tikzpicture}[yscale=-1,transform shape]
\pic{orcidlogo};
\end{tikzpicture}
}{|}}}}
\usepackage{hyperref}

\begin{document}
\definecolor{orcidlogocol}{HTML}{A6CE39}

\title{A System-Level Simulation Module for Multi-UAV IRS-assisted Communications}

\author{
\thanks{This work has been submitted to the IEEE for possible publication. This work was partially supported by the European Union under the Italian National Recovery and Resilience Plan (NRRP) of NextGenerationEU, with particular reference to the partnership on “Telecommunications of the Future” (PE00000001 - program “RESTART”, CUP: D93C22000910001) and to the national center on “Sustainable Mobility” (CN00000023 - program “MOST”, CUP: D93C22000410001). It was also supported by the PRIN project no. 2017NS9FEY entitled “Realtime Control of 5G Wireless Networks: Taming the Complexity of Future Transmission and Computation Challenges“ funded by the Italian MUR, by “The house of emerging technologies of Matera (CTEMT)“ project funded by the Italian MIMIT, and by the PON AGREED projects (ARS01 00254) fundend by the Italian MUR.}
\thanks{G. Grieco, G. Iacovelli, D. Pugliese, D. Striccoli, and L.A. Grieco are with the Department of Electrical and Information Engineering, Politecnico di Bari, Bari, Italy (email: \textit{name.surname}@poliba.it). G. Grieco, G. Iacovelli, D. Striccoli, and L.A. Grieco are with the Consorzio Nazionale Interuniversitario per le Telecomunicazioni, Parma, Italy.}
	\IEEEauthorblockN{
		Giovanni Grieco~\orcidicon{0000-0002-6326-4244},~\IEEEmembership{Graduate Student Member,~IEEE},
        Giovanni Iacovelli~\orcidicon{0000-0002-3551-4584},~\IEEEmembership{Graduate Student Member,~IEEE},
        Daniele Pugliese~\orcidicon{0000-0001-6436-2359},~\IEEEmembership{Student Member,~IEEE},
        Domenico Striccoli~\orcidicon{0000-0003-2904-6961}, and\\
        Luigi Alfredo Grieco~\orcidicon{0000-0002-3443-6924},~\IEEEmembership{Senior Member,~IEEE}\\
	}
}

\acrodef{6G}{Sixth-Generation}

\acrodef{AF}{Amplify-and-Forward}
\acrodef{ATC}{Air Traffic Control}

\acrodef{BS}{Base Station}

\acrodef{eNB}{Evolved Node-B}

\acrodef{FBS}{Flying Base Station}

\acrodef{GSL}{GNU Scientific Library}
\acrodef{GU}{Ground User}
\acrodef{GUI}{Graphical User Interface}

\acrodef{JSON}{JavaScript Object Notation}

\acrodef{KPI}{Key Performance Indicator}

\acrodef{IoD}{Internet of Drones}
\acrodef{IoD-Sim}{Internet of Drones Simulator}
\acrodef{IRS}{Intelligent Reflective Surface}

\acrodef{LoS}{Line of Sight}
\acrodef{LTE}{Long Term Evolution}

\acrodef{MCS}{Modulation and Coding Scheme}

\acrodef{NLoS}{Non Line of Sight}
\acrodef{NR}{New Radio}
\acrodef{ns-3}{Network Simulator 3}

\acrodef{OFDMA}{Orthogonal Frequency Multiple Access}

\acrodef{PRU}{Passive Reflective Unit}

\acrodef{REM}{Radio Environment Map}

\acrodef{SINR}{Signal-to-Interference-plus-Noise Ratio}
\acrodef{SWaP}{Size, Weight, and Power consumption}

\acrodef{UAV}{Unmanned Aerial Vehicle}
\acrodef{UDP}{User Datagram Protocol}
\acrodef{UE}{User Equipment}

\acrodef{ZSP}{Zone Service Provider}

\newcommand{\TODO}{\textcolor{blue}{TODO}}
\newcounter{mytempeqncnt}

\renewcommand{\qedsymbol}{\scalebox{0.75}{$\blacksquare$}}
\newtheorem{assumption}{Assumption}
\newtheorem{remark}{Remark}
\newtheorem{theorem}{Theorem}
\newtheorem{corollary}{Corollary}
\newtheorem{lemma}{Lemma}

\maketitle

\begin{abstract} 
\ac{6G} networks are set to provide reliable, widespread, and ultra-low-latency mobile broadband communications for a variety of industries. In this regard, the \ac{IoD} represents a key component for the development of 3D networks, which envisions the integration of terrestrial and non-terrestrial infrastructures.
The recent employment of \acp{IRS} in combination with \acp{UAV} introduces more degrees of freedom to achieve a flexible and prompt mobile coverage.
As the concept of smart radio environment is gaining momentum across the scientific community, this work proposes an extension module for \ac{IoD-Sim}, a comprehensive simulation platform for the \ac{IoD}, based on \ac{ns-3}. This module is purposefully designed to assess the performance of \ac{UAV}-aided \ac{IRS}-assisted communication systems. Starting from the mathematical formulation of the radio channel, the simulator implements the \ac{IRS} as a peripheral that can be attached to a drone. Such device can be dynamically configured to organize the \ac{IRS} into patches and assign them to assist the communication between two nodes. Furthermore, the extension relies on the configuration facilities of \ac{IoD-Sim}, which greatly eases design and coding of scenarios in \ac{JSON} language. A simulation campaign is conducted to demonstrate the effectiveness of the proposal by discussing several \acp{KPI}, such as \ac{REM}, \ac{SINR}, maximum achievable rate, and average throughput.
\end{abstract}


\acresetall

\section{Introduction}\label{sec:introduction}
\ac{6G} networks promise reliable and ubiquitous mobile broadband, as well as massive ultra-low-latency communications. These characteristics answer the emerging needs of manifold verticals, such as eHealth, intelligent transport systems, immersive multimedia entertainment, automotive, and cyber-physical security \cite{9040264,9714139,9914805}.

In this context, the \ac{IoD} \cite{BSG21} is paving the way to 3D networks, where classical terrestrial and non-terrestrial infrastructure are integrated to provide connectivity in harsh environments, including oceans, deserts, and hazardous places \cite{9861699}. Indeed, \acp{UAV} play a central role in the design of future communication technologies, as they offer high mobility for an on-demand network coverage, i.e., \acp{FBS} \cite{9734063}.

One of the most challenging aspects that these systems encounter is the Shannon capacity limit, which is especially bounded by the available bandwidth. For this reason, the research and standardization communities are focusing on mmWave and THz spectrum to unlock ultra-wide channel capacity \cite{9768113,6016195,10044617}. Nonetheless, the environment can also be controlled to turn adverse effects, such as multipath, into advantages. In this regard, \acp{IRS} \cite{9122596} allow to control the radio environment by optimally reflecting incident electromagnetic waves through a matrix of \acp{PRU}, thus yielding passive beamforming \cite{ICG21}.

Differently from the traditional antenna array systems, \acp{IRS} can not only be deployed as fixed, standalone entities on buildings, but they also satisfy \ac{SWaP} constraints required by drones. Consequently, the integration of \acp{IRS} and \acp{UAV} leads to more degrees of freedom that can be properly tuned to cope with the ever-changing channel, providing the possibility to re-establish the \ac{LoS} and to reduce the pathloss \cite{ICG21}.

Although this novel communication infrastructure paradigm is quite compelling, to the best of authors' knowledge, the scientific literature \cite{HFSR22,CC21,BY2020,PGG+22,SPR22,GWS+22} did not consider the presence of drones, as it proposes solutions solely focused on \acp{IRS}-aided communication systems.
In this regard, \cite{CC21} introduces WiThRay, a versatile framework which models the mmWave channel response in 3D environments by employing ray tracing. It allows to deploy and configure multiple \acp{BS} and \acp{IRS}, which serve mobile users.
In \cite{BY2020}, an open-source MATLAB-based simulator is developed, namely SimRIS, which leverages a channel model for mmWave frequencies, applicable in various indoor and outdoor environments. The simulator provides a simple \ac{GUI} which gives the possibility to set up (i) the operating frequency, (ii) the terminal locations, and (iii) the number of IRS elements.
\cite{PGG+22} proposes a simulation framework, based on \ac{ns-3}, to simulate \ac{IRS} and \ac{AF} systems. The end-to-end communication is implemented by employing the standardized 3GPP TR 38.901 channel model \cite{3gpp.38.901} and the 5G \ac{NR} protocol stack. This contribution aims to (i) demonstrate whether \ac{IRS}/\ac{AF} nodes can be used to relay network traffic and (ii) dimension the number of \ac{IRS}/\ac{AF} nodes with respect to the number of users.
\cite{SPR22} analyzes the system-level simulation results of urban scenarios in which multiple \ac{IRS} are deployed in presence of a 5G cellular network. It emerges that the \ac{IRS} performance strongly depends on its size and the operating frequency. In particular, this manuscript investigates the benefits brought by \acp{IRS}, in
mid (C-band) and high (mmWave) frequency bands, by deriving outdoor and indoor coverage and per-resource block rate.
Similarly to previous works, \cite{GWS+22} introduces a system-level simulation platform implemented in C++ for 5G systems, which includes different features, such as network topology, antenna pattern, large/small scale channel models, and many performance
indicators. Specifically, this paper investigates the case in which the \ac{LoS} propagation is dominant under far-field conditions. Moreover, the performance of phase quantization are also discussed and analyzed. Besides, \cite{HFSR22} implements an
extension for the Vienna 5G simulator, which includes \ac{IRS} modeling, \ac{IRS} phase shifts optimization, large- and small-scale fading.

The contributions discussed above consider each surface associated only to a specific user that, on one hand, simplifies the mathematical modeling and the software implementation, but, on the other hand, limits the achievable system performance.
Furthermore, the employment of aerial mobile \acp{IRS}, enabled by drones, is not taken into account, even if it would (i) represent a big advantage in terms of flexibility and (ii) increase the scenario complexity.

In light of the above, the major contributions given by this work are listed below.
\begin{itemize}[topsep=0pt]
    \item A channel model expression for \ac{UAV}-aided \ac{IRS}-assisted communications is derived. In particular, a swarm of \ac{IRS}-equipped drones is considered, in charge of enhancing the channel quality of \acp{GU}. The system adopts the \ac{OFDMA} scheme, which avoids interference among users. Nonetheless, the mathematical formulation still considers constructive/destructive interference patterns due to the presence of multiple \acp{IRS}. Further, the \acp{IRS} are divided into patches of an arbitrary size, which can be assigned to different \acp{GU}. Based on these assumptions, a gain lowerbound expression is obtained by (i) reducing the number of degrees of freedom introduced by the controllable phase shifts, (ii) employing a mathematical approximation for the complex gaussian product involved in the channel modeling, and (iii) imposing a fixed outage probability to cope with the inherent stochasticity of the channel.
    \item On top of \ac{IoD-Sim} \cite{iodsim}, an \ac{IRS} simulation module, based on the previously derived channel model, is proposed\footnote{The source code is freely available at the following URL: \url{https://telematics.poliba.it/iod-sim}}. Its architecture and the consequent implementation are deeply discussed, as well as all the developed functionalities. Indeed, it is possible to configure the whole mission by properly set up a \ac{JSON} file. Among the manifold configuration parameters, the proposed module provides the possibility to dynamically change (i) the number and the size of the patches, and (ii) for how long a certain \ac{GU} is served by a specific patch. Moreover, thanks to the fact that \ac{IoD-Sim} is based on \ac{ns-3}, it is possible to employ an arbitrary communication stack on top of the PHY layer provided by the module.
    \item A simulation campaign is carried out to prove the validity of this work. To this end, three different scenarios are investigated under different configuration settings by taking into account several \acp{KPI}, such as \ac{REM}, \ac{SINR}, maximum achievable rate, and average throughput.
  \end{itemize}
The numerical results obtained from the proposed \ac{IRS} simulator indicate that the presence of \ac{IRS}-equipped drones enhances the channel quality of the \acp{GU}. Moreover, the possibility to organize the \ac{IRS} in patches is an effective solution to uniformly assist multiple nodes. This in turn demonstrates the unique potential of the simulation platform to assess and prototype complex \ac{IoD}-enabled \acp{IRS} systems.

The remainder of the present contribution is as follows: Section \ref{sec:sysmod} describes the adopted system model. Section \ref{sec:chmod} presents the proposed channel model.
Section \ref{sec:swdesign} describes the proposed solution, i.e., the \ac{IRS} module, integrated with \ac{IoD-Sim}. Section \ref{sec:simcampaign} analyzes accuracy of the model and investigates the obtained numerical results.
Finally, Section \ref{sec:conclusions} concludes the work and draws future research perspectives.

Notations adopted in this work are hereby described. Boldface lower and capital case letters refer to vectors and matrices, respectively; $j = \sqrt{-1}$ is the imaginary unit; $\text{atan2}\left(x\right)$ denotes the four-quadrant arctangent of a real number $x$;
$\textbf{x}^\mathsf{T}$ is the transpose of a generic vector $\textbf{x}$;
$\text{diag}(\textbf{x})$ represents a diagonal matrix whose diagonal is given by a vector $\textbf{x}$.
For clarity, the adopted notations of this paper are summarized in Table \ref{tab:notation}.

\begin{table}[ht]
\centering
\scriptsize
\begin{tabular}{|l|l|l|l|}
\hline
\multicolumn{1}{|c|}{\textbf{Symbol}} & \multicolumn{1}{c|}{\textbf{Description}} & \multicolumn{1}{c|}{\textbf{Symbol}} & \multicolumn{1}{c|}{\textbf{Description}} \\ \hline
$K$ & Mission duration. & $d^\text{\tiny{BG}}$ & \acs{BS}-\acs{GU} distance. \\ \hline
$U$ & Number of \acsp{UAV}. & $d^\text{\tiny{UG}}_{u}$ & \acs{UAV}-\acs{GU} distance. \\ \hline
$G$ & Number of \acsp{GU}. & $d^\text{\tiny{BU}}_{u}$ & \acs{BS}-\acs{UAV} distance. \\ \hline
$M^\text{\tiny{R}}$ & \acsp{PRU} as patch rows. & $g^\text{\tiny{BG}}$ & \acs{GU}-\acs{BS} direct link gain. \\ \hline
$M^\text{\tiny{C}}$ & \acsp{PRU} as patch columns. & $\beta^\text{\tiny{BG}}$ & \acs{BS}-\acs{GU} power gain at $\SI{1}{m}$. \\ \hline
$\textbf{q}^\text{\tiny{BS}}$ & Location of the BS. & $\alpha$ & \acs{BS}-\acs{GU} link pathloss exponent. \\ \hline
$\textbf{q}^\text{\tiny{G}}$ & Location of the \acsp{GU}. & $\kappa^\text{\tiny{BG}}$ & K-factor for \acs{BS}-\acs{GU} link. \\ \hline
$\textbf{q}_{u}^\text{\tiny{U}}$ & $u$-th \acs{UAV} location. & $\Omega^\text{\tiny{BG}}$ & \acs{BS}-\acs{GU} link average power. \\ \hline
$v_{u}$ & $u$-th \acs{UAV} speed. & $\textbf{g}_{u,p}^\text{\tiny{UG}}$ & Patch-\acs{GU} channel gain. \\ \hline
$\phi_{u,p,m}$ & $m$-th \acs{PRU} phase shift. & $\textbf{g}_{u,p}^\text{\tiny{BU}}$ & Patch-\acs{BS} channel gain. \\ \hline
$f$ & The carrier frequency. & $\mathbf{\Phi}_{u,p}$ & Phase shift matrix. \\ \hline
$w$ & \acs{PRU} area. & $P$ & Number of \acs{IRS} patches. \\ \hline
$\kappa_{u}^\text{\tiny{UG}}$ & K-factor for UAV-\acs{GU} link. & $\kappa_{u}^\text{\tiny{BU}}$ & K-factor for \acs{BS}-\acs{UAV} link. \\ \hline
\end{tabular}
\caption{Main notation adopted in this work.}
\label{tab:notation}
\end{table}
\begin{figure*}[!t]
    \centering
    \includegraphics[width=0.8\textwidth]{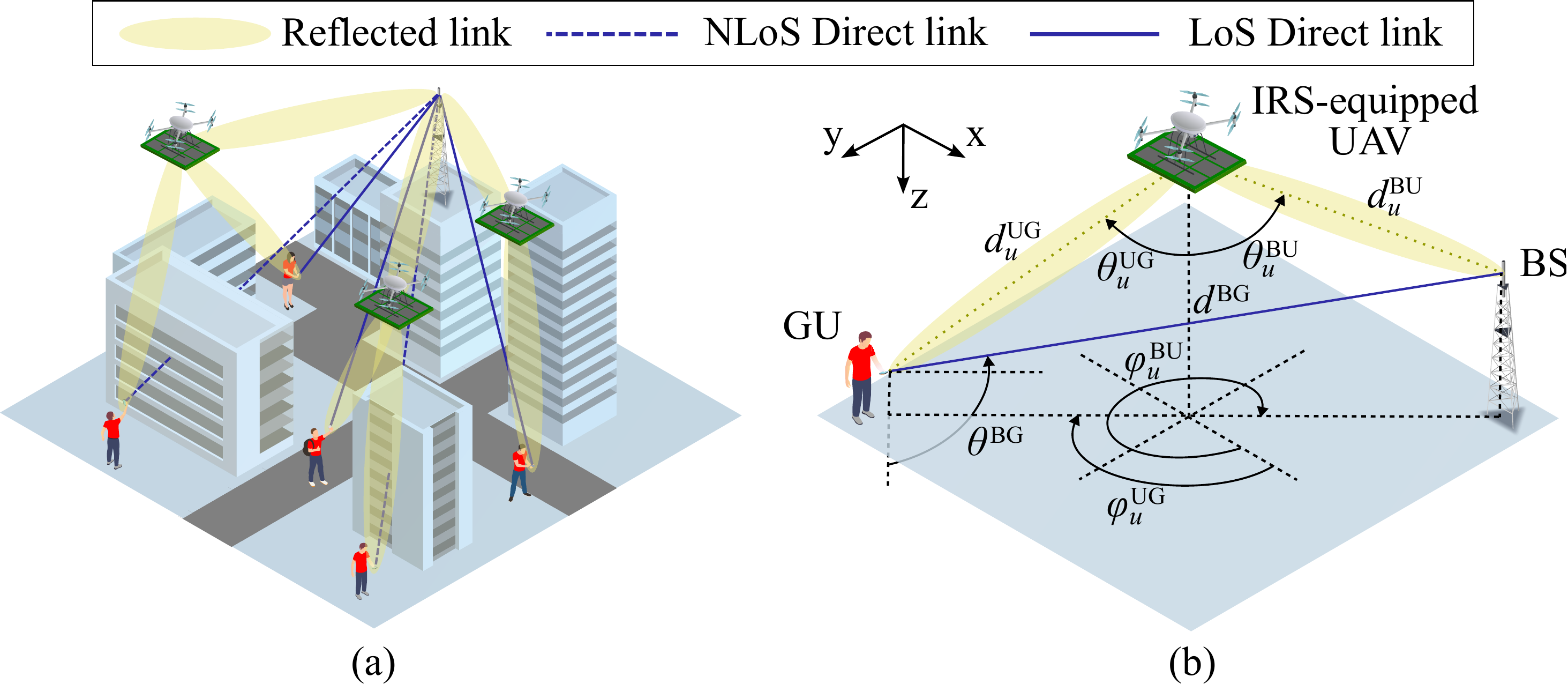}
    \caption{Overview of (a) the reference scenario and (b) the channel model geometry.}
    \label{fig:scenario}
\end{figure*}

\section{System Model}\label{sec:sysmod}
The investigated scenario, illustrated in Figure \ref{fig:scenario}, considers a mission that lasts $K$ seconds, in which $U$ \ac{IRS}-equipped \acp{UAV}, flying at speed $v_{u} \in \mathbb{R}$, $u=1,\ldots,U$, improve the channel quality between a set of $G$ \acp{GU} and the \ac{BS}, through a proper signal reflection. In this Section, for the sake of notations, the discussion considers the communication of a \ac{GU}, in a given subcarrier centered in $f \ \si{Hz}$, at a certain time instant\footnote{It is assumed that the Doppler effect is perfectly compensated, since the parameters related to the kinetics are known to the \ac{BS}.}. The positions of the drones, the \ac{GU}, and the \ac{BS} are denoted as $\textbf{q}_{u}^\text{\tiny{U}} = [x^\text{\tiny{U}}_{u}, y^\text{\tiny{U}}_{u}, z_{u}^\text{\tiny{U}}]^\mathsf{T} \in \mathbb{R}^3$, $\textbf{q}^\text{\tiny{G}} = [x^\text{\tiny{G}}, y^\text{\tiny{G}}, z^\text{\tiny{G}}]^\mathsf{T}\in \mathbb{R}^3$, and $\textbf{q}^\text{\tiny{B}} = [x^\text{\tiny{B}}, y^\text{\tiny{B}}, z^\text{\tiny{B}}]^\mathsf{T} \in \mathbb{R}^3$.
Accordingly, the far-field distances $d^\text{\tiny{UG}}_{u}$, $d^\text{\tiny{BU}}_{u}$, and $d^\text{\tiny{BG}}$ are defined as $d^\text{\tiny{ab}} = \norm{\textbf{q}^\text{\tiny{a}} - \textbf{q}^\text{\tiny{b}}}$, with a, b~$\in \{U,G,B\}$.

\acp{IRS} are composed by $N = N^\text{\tiny{R}}\times N^\text{\tiny{C}}$ \acp{PRU}, having the size $w = d^\text{\tiny{X}} \times d^\text{\tiny{Y}}\ \si{m^2}$, with $d^\text{\tiny{X}} = d^\text{\tiny{Y}} \triangleq d$ being the length of the element sides.
The midpoint of each \ac{PRU}, with respect to the center of the \ac{IRS}, is $d\left[(i-\frac{1}{2}), (i^\prime-\frac{1}{2})\right]^\mathsf{T}$ with $i=1-\frac{N^\text{\tiny{C}}}{2},\ldots,\frac{N^\text{\tiny{C}}}{2}, i^\prime=1-\frac{N^\text{\tiny{R}}}{2},\ldots,\frac{N^\text{\tiny{R}}}{2}$.
The \acp{PRU} are grouped into $P$ patches of $M = M^\text{\tiny{R}} \times M^\text{\tiny{C}}$ elements, each one indexed as $m=1,\ldots,M$. Moreover, each patch reflects the incident signal according to a phase shift matrix $\mathbf{\Phi}_{u,p} \in \mathbb{C}^{M\times M}$, with $p = 1,\ldots,P$, defined as
\begin{align}
    \mathbf{\Phi}_{u,p} = \text{diag}\bigg(&e^{j\phi_{u,p,1}},\ldots, e^{j\phi_{u,p,m}}, \ldots, e^{j\phi_{u,p,M}}\bigg),
    \label{eq:phasematrix}
\end{align}
where $\phi_{u,p,m} \in [-\pi, \pi)$. It is worth specifying that for ease of readability, all the \acp{IRS} have the same number and size of patches but the model is straightforward extensible. This is demonstrated by the actual implementation of the simulator, described in the Section \ref{sec:swdes}.

Finally, define $\{\theta^\text{\tiny{UG}}_{u},\theta^\text{\tiny{BU}}_{u}\}$ and $\{\varphi^\text{\tiny{UG}}_{u},\varphi^\text{\tiny{BU}}_{u}\}$ as the inclination and azimuth angles between the center of the \ac{IRS} and the \ac{GU}/\ac{BS} as
$\theta^\text{\tiny{ab}} = \mathrm{acos}\left(\frac{z^\text{\tiny{a}}-z^\text{\tiny{b}}}{d^\text{\tiny{ab}}}\right)$ and $\varphi^\text{\tiny{ab}} = \mathrm{atan2}\left(\frac{y^\text{\tiny{a}}-y^\text{\tiny{b}}}{x^\text{\tiny{a}}-x^\text{\tiny{b}}}\right)$.
Similarly, $\theta^\text{\tiny{BG}}$ denotes the inclination angle related to the direct \ac{GU}-\ac{BS} link with respect to the \ac{GU}.

\section{Channel Model}\label{sec:chmod}
The communication system employs the \ac{OFDMA} scheme, which prevents interference among the involved entities.
The \ac{GU} and the \ac{BS} employ a single-antenna for data exchange, that, together with each \ac{IRS} element, are characterized by power radiation pattern functions (including antenna gains) denoted by $F^\text{\tiny{GU}}$, $F^\text{\tiny{BS}}$, and $F^\text{\tiny{IRS}}$.

According to \cite{Iacovelli2023}, the channel gain $g^\text{\tiny{BG}} \in \mathbb{C}$ of the direct \ac{GU}-\ac{BS} link is
\begin{align}
    &g^\text{\tiny{BG}} = \sqrt{\beta^\text{\tiny{BG}} {d^\text{\tiny{BG}}}^{-\alpha}F^\text{\tiny{BG}}}h^\text{\tiny{BG}},\label{eq:gbg}
\end{align}
where $\beta^\text{\tiny{BG}}$ is the channel power gain at the reference distance of 1 m, $\alpha$ is the pathloss exponent, and $F^\text{\tiny{BG}} = F^\text{\tiny{BS}}F^\text{\tiny{GU}}$. Moreover, $h^\text{\tiny{BG}}$ is the channel coefficient, which accounts for the small-scale fading and follows a circularly-symmetric non-central complex gaussian distribution. The envelope $|h^\text{\tiny{BG}}|$ is generally Rician \cite{APA+19}, with K-factor $\kappa^\text{\tiny{BG}}$ and average power $\Omega^\text{\tiny{BG}} = 1$. Specifically, $\kappa^\text{\tiny{BG}}$ can be expressed as a function of the elevation angle and reads
\begin{equation}
    \kappa^\text{\tiny{BG}} = \kappa^\text{\tiny{MIN}}\exp\left(\frac{2}{\pi}\ln{\frac{\kappa^\text{\tiny{MAX}}}{\kappa^\text{\tiny{MIN}}}}\left|\frac{\pi}{2}-\theta^\text{\tiny{BG}}\right|\right),
\end{equation}
with $\kappa^\text{\tiny{MIN}}$ and $\kappa^\text{\tiny{MAX}}$ the minimum and maximum possible K-factors, respectively.
With similar definitions, given the $p$-th patch of the $u$-th \ac{UAV}, the channel gains $\textbf{g}_{u,p}^\text{\tiny{UG}} \in \mathbb{C}^M$ and $\textbf{g}_{u,p}^\text{\tiny{BU}} \in \mathbb{C}^M$, related to the \ac{GU} and the \ac{BS}, can be formulated as follows:
\begin{align}
    &\textbf{g}_{u,p}^\text{\tiny{UG}} = \sqrt{\beta^\text{\tiny{UG}} {d^\text{\tiny{UG}}_{u}}^{-2} F^\text{\tiny{UG}}} \textbf{h}_{u,p}^\text{\tiny{UG}},\label{eq:gug}\\
    &\textbf{g}_{u,p}^\text{\tiny{BU}} = \sqrt{\beta^\text{\tiny{BU}} {d^\text{\tiny{BU}}_{u}}^{-2} F^\text{\tiny{BU}}} \textbf{h}_{u,p}^\text{\tiny{BU}}\label{eq:gbu},
\end{align}
whose envelopes are characterized by K-factors $\kappa_{u}^\text{\tiny{UG}}$ and $\kappa_{u}^\text{\tiny{BU}}$.
Since each patch $p$ coherently reflects the incident signal from the \ac{BS} towards a \ac{GU} and vice versa, all the phase shifts can be described in terms of two parameters, $\phi_{u,p}^\text{\tiny{X}}$ and $\phi_{u,p}^\text{\tiny{Y}}$, thus reducing the degrees of freedom by imposing that:
\begin{align}
    &\ell\left(\left(i-\frac{1}{2}\right)\phi_{u,p}^\text{\tiny{X}}+\left(i^\prime-\frac{1}{2}\right)\phi_{u,p}^\text{\tiny{Y}}\right) = \phi_{u,p,m},
\end{align}
being $\ell = \frac{2\pi f d}{c}$ and $c$ the speed of light.
The overall channel gain that characterizes the communication of a \ac{GU} served by the swarm is
\begin{align}
    \Gamma = \sum_{u=1}^{U}\sum_{p=1}^{P}{\textbf{g}^\text{\tiny{BU}}_{u,p}}^\mathsf{T}\mathbf{\Phi}_{u,p}\textbf{g}_{u,p}^\text{\tiny{UG}} + g^\text{\tiny{BG}} \label{eq:G},
\end{align}
which is intractable due to the product of complex gaussians. Nonetheless, according to \cite{Iacovelli2023}, the envelope can be approximated to a Rician random variable having K-factor $\kappa = \frac{\nu^2}{2{\sigma}^2}$ and average power $\Omega = \nu^2+2{\sigma}^2$, with $\nu^2$ and $2{\sigma}^2$ defined as
\begin{align}
    \nu^2 &= \sum^{U}_{u=1}\sum^{P}_{p=1}\mu_{u,p}^2 + 2\sum_{u\geq u'}\sum_{p>p'}|\mu_{u,p}||\mu_{u',p'}| \cos\left(\omega_u -\omega_{u'}\right)\nonumber\\&+\!\lambda^2\overline{\kappa}^\text{\tiny{BG}}\!+\! 2\sum^{U}_{u=1}\sum^{P}_{p=1}|\mu_{u,p}||\lambda\sqrt{\overline{\kappa}^\text{\tiny{BG}}}|\cos\left(\omega_u\!+\!\frac{\ell d_g^\text{\tiny{BG}}}{d}\right),\label{eq:nug}\\
    2\sigma^2 &= N\sum_{u=1}^{U}\eta^2_{u}\widetilde{\kappa}^{\text{\tiny{BUG}}}_{u} + \lambda^2 \widetilde{\kappa}^{\text{\tiny{BG}}},\label{eq:sigmag}
\end{align}
with
\begin{align}
    &\mu_{u,p} = \eta_{u}\sqrt{\overline{\kappa}_{u}^\text{\tiny{BUG}}} \frac{\sin\left(\frac{\ell M^\text{\tiny{C}}}{2}\psi^\text{\tiny{X}}_{u,p}\right)\sin\left(\frac{\ell M^\text{\tiny{R}}}{2}\psi^\text{\tiny{Y}}_{u,p}\right)}{\sin\left(\frac{\ell}{2}\psi^\text{\tiny{X}}_{u,p}\right)\sin\left(\frac{\ell}{2}\psi^\text{\tiny{Y}}_{u,p}\right)}e^{-j\omega_u},\label{eq:muGp}\\
    &\overline{\kappa}_{u}^\text{\tiny{BUG}} = \frac{\kappa_{u}^\text{\tiny{BU}}\kappa_{u}^\text{\tiny{UG}}}{(\kappa_{u}^\text{\tiny{BU}} +1)(\kappa_{u}^\text{\tiny{UG}}+1)},
    \widetilde{\kappa}_{u}^\text{\tiny{BUG}} =\frac{\kappa_{u}^\text{\tiny{BU}}+\kappa_{u}^\text{\tiny{UG}}}{(\kappa_{u}^\text{\tiny{BU}} +1)(\kappa_{u}^\text{\tiny{UG}}+1)},\\&
    \psi^\text{\tiny{X}}_{u,p} = \sin{\theta^\text{\tiny{BU}}_{u}}\cos{\varphi^\text{\tiny{BU}}_{u}}+\sin{\theta^\text{\tiny{UG}}_{u}}\cos{\varphi^\text{\tiny{UG}}_{u}} + \phi_{u,p}^\text{\tiny{X}},\\&
    \psi^\text{\tiny{Y}}_{u,p} = \sin{\theta^\text{\tiny{BU}}_{u}}\sin{\varphi^\text{\tiny{BU}}_{u}}+\sin{\theta^\text{\tiny{UG}}_{u}}\sin{\varphi^\text{\tiny{UG}}_{u}} + \phi_{u,p}^\text{\tiny{Y}},
\end{align}
$\omega_u = \ell\left(d^\text{\tiny{BU}}_{u}+d^\text{\tiny{UG}}_{u}\right),\qquad\eta_{u} = \sqrt{\beta^\text{\tiny{BUG}}{d^\text{\tiny{BU}}_{u}}^{-2}{d^\text{\tiny{UG}}_{u}}^{-2} F^\text{\tiny{BUG}}}$, $\lambda~=~\sqrt{\beta^\text{\tiny{BG}} {d^\text{\tiny{BG}}}^{-\alpha}F^\text{\tiny{BG}})}$, $F^\text{\tiny{BUG}} = F^\text{\tiny{BU}}F^\text{\tiny{UG}}$, $\overline{\kappa}^{\text{\tiny{BG}}} = \frac{\kappa^{\text{\tiny{BG}}}}{\kappa^{\text{\tiny{BG}}} + 1}$, $\widetilde{\kappa}^{\text{\tiny{BG}}} = \left(\kappa^\text{\tiny{BG}}+1\right)^{-1}$, and $\beta^\text{\tiny{BUG}} = \beta^\text{\tiny{BU}}\beta^\text{\tiny{UG}}$.

Finally, given an outage probability $\varepsilon$, the channel power gain can be lowerbounded \cite{Iacovelli2023} as
\begin{align}
    \Gamma_\varepsilon &= \frac{\zeta^2\Omega}{2(\kappa+1)},\label{eq:gain}\\
  \zeta&=  \begin{cases} \label{eq:gain2}
    \sqrt{-2\log(1-\varepsilon)}e^{\frac{\kappa}{2}}, & \! \! \! \! \!\text{for } \kappa \leq \frac{K_0^2}{2}\\
    \sqrt{2\kappa}+\frac{1}{2Q^{-1}(\varepsilon)} \times\\
    \quad\log\left(\frac{\sqrt{2\kappa}}{\sqrt{2\kappa} - Q^{-1}(\varepsilon)}\right)-Q^{-1}(\varepsilon), & \! \! \! \! \!\text{for } \kappa > \frac{K_0^2}{2}
\end{cases}
\end{align}
with $Q^{-1}(x)$ being the inverse Q-function and $K_0$ the intersection of the sub-functions at $\sqrt{2\kappa} > Q^{-1}(\varepsilon)$. Equations \eqref{eq:gain} and \eqref{eq:gain2} will be used in the proposed \ac{IoD-Sim} \ac{IRS} module.

\section{Software Design}\label{sec:swdesign}
The proposed module is designed according to the structure of \ac{IoD-Sim}. To this end, a general overview of the simulator architecture is given, thus providing a comprehensive description of the abstraction layers and the available software facilities. Then, the \ac{IRS} module is introduced, along with the software definition and the mathematical model implementation. Finally, the configuration of a simple scenario, via \ac{JSON} parameters, is discussed.

\subsection{IoD Sim}
\begin{figure}[!ht]
    \centering
    \includegraphics[width=\columnwidth]{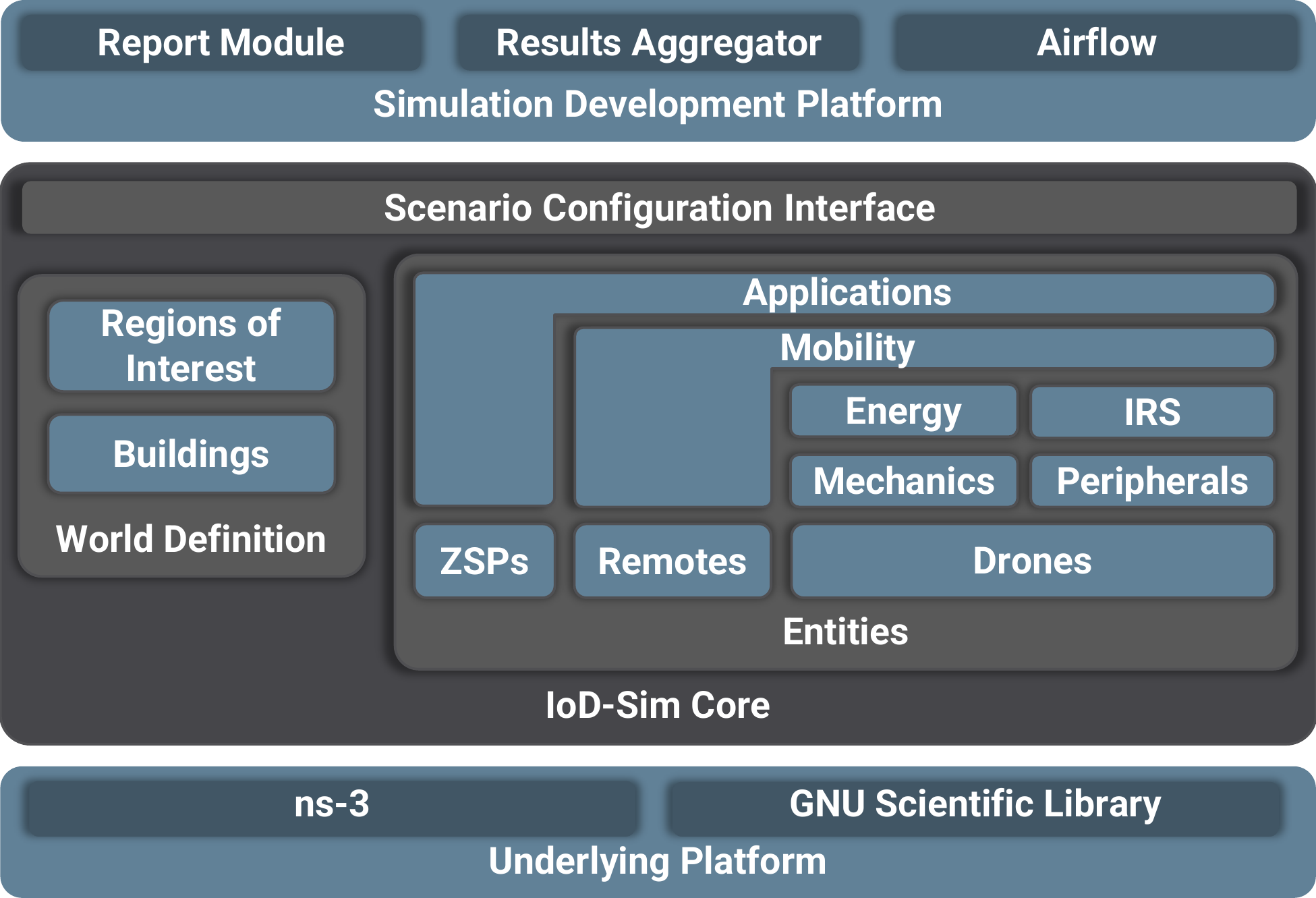}
    \caption{\ac{IoD-Sim} architecture integrated with the \ac{IRS} module.}
    \label{fig:iodsim_arch}
\end{figure}
\begin{figure*}[!ht]
    \centering
    \includegraphics[width=\textwidth]{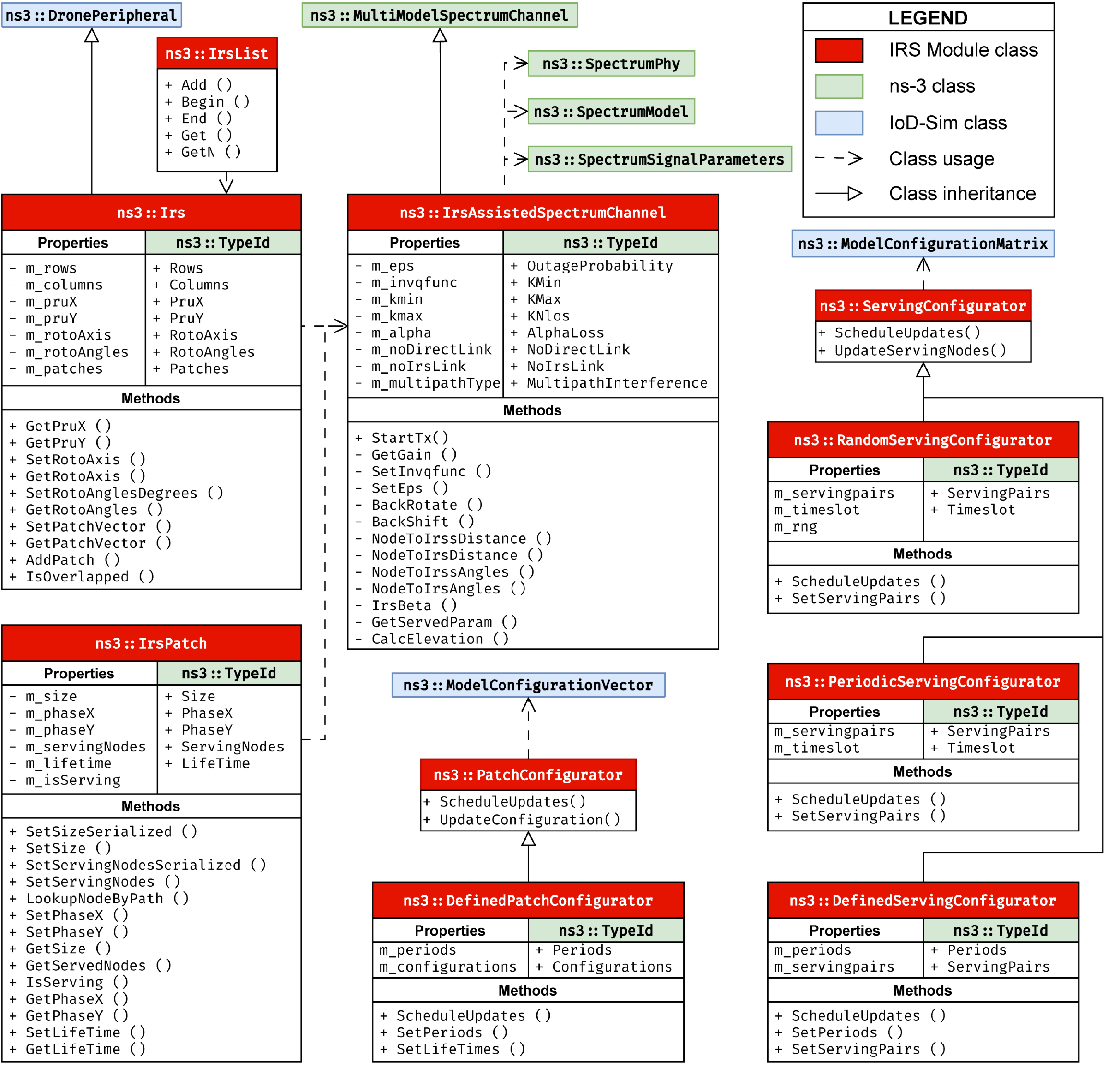}
    \caption{Class diagram of the \ac{IRS} extension implemented in \ac{IoD-Sim}.}
    \label{fig:class_diagram}
\end{figure*}

\ac{IoD-Sim} \cite{iodsim} is a comprehensive simulation platform designed to assess \ac{UAV}-enabled communication systems with ease.
The architecture, illustrated in Figure \ref{fig:iodsim_arch}, is a software stack composed by three logical layers, namely \textit{Underlying Platform}, \textit{\ac{IoD-Sim} Core}, and \textit{Simulation Development Platform}.
The former provides a solid foundation of established software used for (i) network simulation, i.e., \ac{ns-3} \cite{Riley2010}, and (ii) optimized mathematical computation, i.e., \ac{GSL}.

On top of that, the simulator presents \textit{\ac{IoD-Sim} Core} which mainly (i) introduces \ac{IoD} entities, (ii) defines the reference 3D world, and (iii) simulates remote cloud services.
In particular, drones are mechanically modeled by taking into account physical and power-consumption properties, such as mass, rotor disk area, drag coefficient of the rotor blades, and battery model. Moreover, \ac{UAV} motion can be simulated through one of the manifold mobility models, which easily allow trajectory design starting from a set of points of interest. Further, applications and peripherals enable the simulation of multiple use cases, spanning from telemetry reporting to multi-stack relaying, video recording, and streaming. To provide services that go beyond classical network coverage, \acp{ZSP} are also implemented in the simulator, which are specialized \acp{BS} that leverages the cyber-physical environment to provide local \ac{ATC}, weather forecast, and cloud services to a geographical zone of interest. On these premises, the platform enables the design of a fully integrated terrestrial/non-terrestrial drone network.

All the features that characterize the \textit{\ac{IoD-Sim} Core} are made available through a low-code and user-friendly \textit{Simulation Development Platform}, which is conceived to easily create and maintain a scenario configuration through either a block-based \ac{GUI}, named Airflow, or a \ac{JSON} file. These designs are then parsed and executed, without requiring deep understanding of the underlying C++ code.

Finally, the software generates simulation reports in the form of plain text and structured data sets that can be processed through conventional data-analysis tools. These may be used to evaluate and graphically analyze \acp{UAV} trajectories, network traffic, and application-specific \acp{KPI}.

\subsection{\ac{IRS} Simulation Module}\label{sec:swdes}

The proposed module is implemented on top of \ac{IoD-Sim}, which represents a solid foundation to design and assess the desired \ac{IRS}-assisted \ac{UAV}-aided scenarios.
Accordingly, multiple classes, depicted in Figure \ref{fig:class_diagram}, are hereby introduced to enhance the \textit{\ac{IoD-Sim} Core}.

In particular, PHY layer communications are implemented by means of the \texttt{ns3::IrsAssistedSpectrumChannel} class, which extends the channel simulation capabilities originally provided by \texttt{ns3::MultiModelSpectrumChannel}. Specifically, this object evaluates the overall receiver gain\footnote{It is worth specifying that, in order to minimize the computational complexity, the gain is calculated only in the center frequency of the signal power spectrum density, instead of iterating over each spectrum component. Since the bandwidth used by each user is much smaller than the carrier frequency, this approximation leads to a negligible frequency shift and hence an accurate channel gain evaluation.}, derived in Section \ref{sec:chmod}, which considers both the reflected links, introduced by the \acp{IRS}, and the original direct link between the nodes of interest.

\begin{figure}[!ht]
    \centering
    \includegraphics[width=\columnwidth]{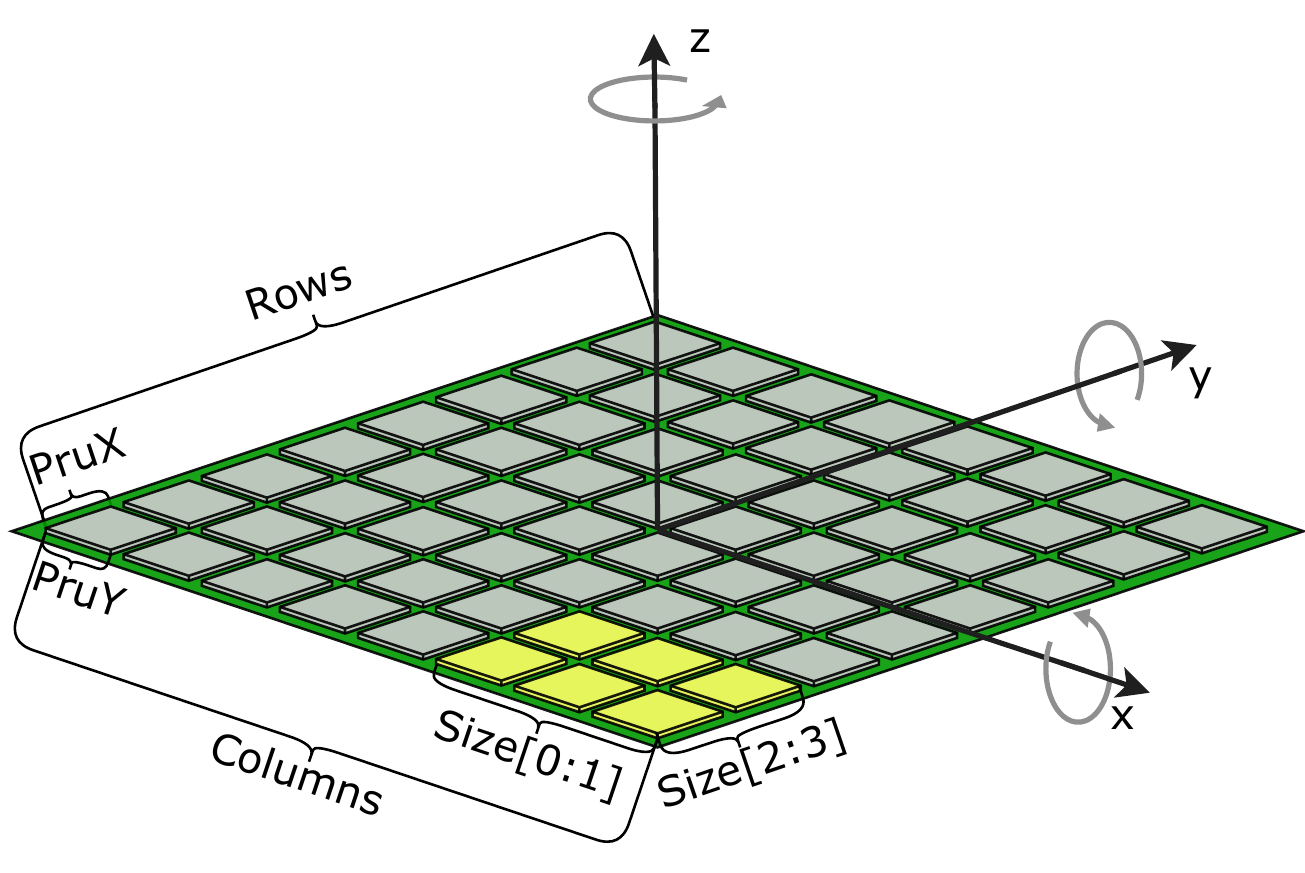}
    \caption{JSON configuration properties of an \ac{IRS} with a patch highlighted in yellow.}
    \label{fig:irs}
\end{figure}
\begin{figure}[!ht]
    \centering
    \includegraphics[width=\columnwidth]{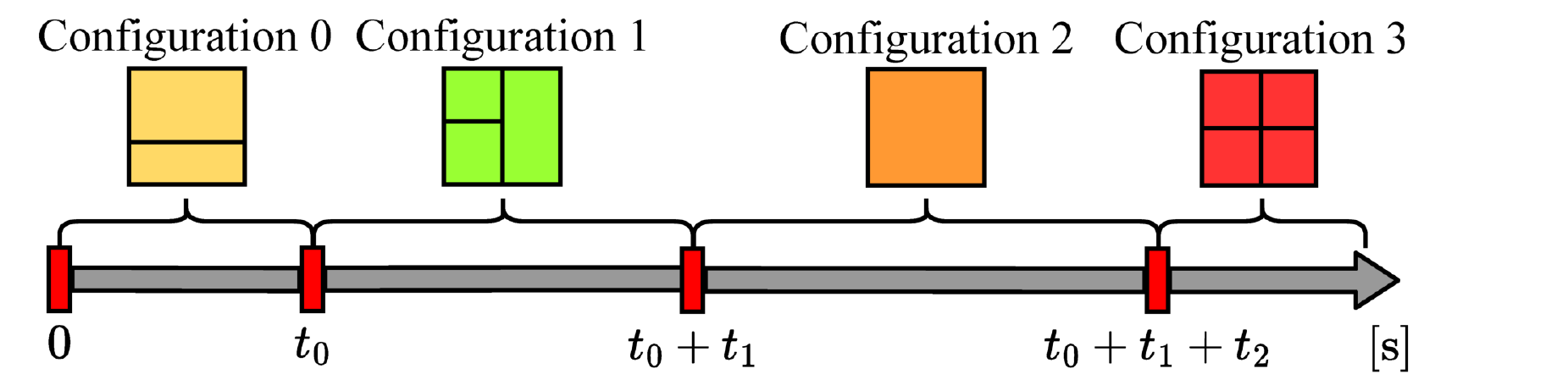}
    \caption{Different patch configurations applied over time, following the \texttt{ns3::DefinedPatchConfigurator} logic.}
    \label{fig:defined_patch_configurator}
\end{figure}

The \ac{IRS} is described by the \texttt{ns3::Irs} class, which extends the generic peripheral one, i.e., \texttt{ns3::DronePeripheral} \cite[Sec. V.B]{iodsim}. The adoption of this interface benefits the implementation of the \ac{IRS} as a device. In fact, it is possible to have (i) a state that can be put in either \texttt{OFF}, \texttt{IDLE}, or \texttt{ON}, and (ii) an associated energy consumption model (even though it is negligible with respect to the main components that drain the \ac{UAV} battery).
All \texttt{ns3::Irs} instances are referenced by a global register named \texttt{ns3::IrsList}, allowing them to be easily reachable through object paths formatted as \texttt{/IrsList/[IRS\_Global\_Index]}.

As illustrated in Figure \ref{fig:irs}, the \ac{IRS} is characterized by different properties that can be set through the \texttt{ns3::TypeId} attributes: \textit{Rows} and \textit{Columns} for the \ac{IRS} size; \texttt{PruX} and \texttt{PruY} for the dimension of each \ac{PRU}; \texttt{RotoAxis} and \texttt{RotoAngles} to indicate an ordered sequence of axes and their rotation in degrees, respectively.
For instance, \texttt{RotoAxis = ["X\_AXIS"]} and \texttt{RotoAngles = [180]} indicate that the \ac{IRS} should be rotated by 180 degrees around the x axis, i.e., the surface faces the ground.

Each \texttt{ns3::Irs} is organized into one or more \texttt{ns3::IrsPatch}, whose dimensions can be specified through the \texttt{Size} property. This property has four values corresponding to the starting and ending \acp{PRU}' indexes along the x and y axes, i.e, \texttt{Size[0:1]} and \texttt{Size[2:3]}.
Once the patches dimensions are set, they can be configured to support the communication of a specific pair of \textit{Serving Nodes}.

\begin{figure}[!ht]
    \centering
    \includegraphics[width=\columnwidth]{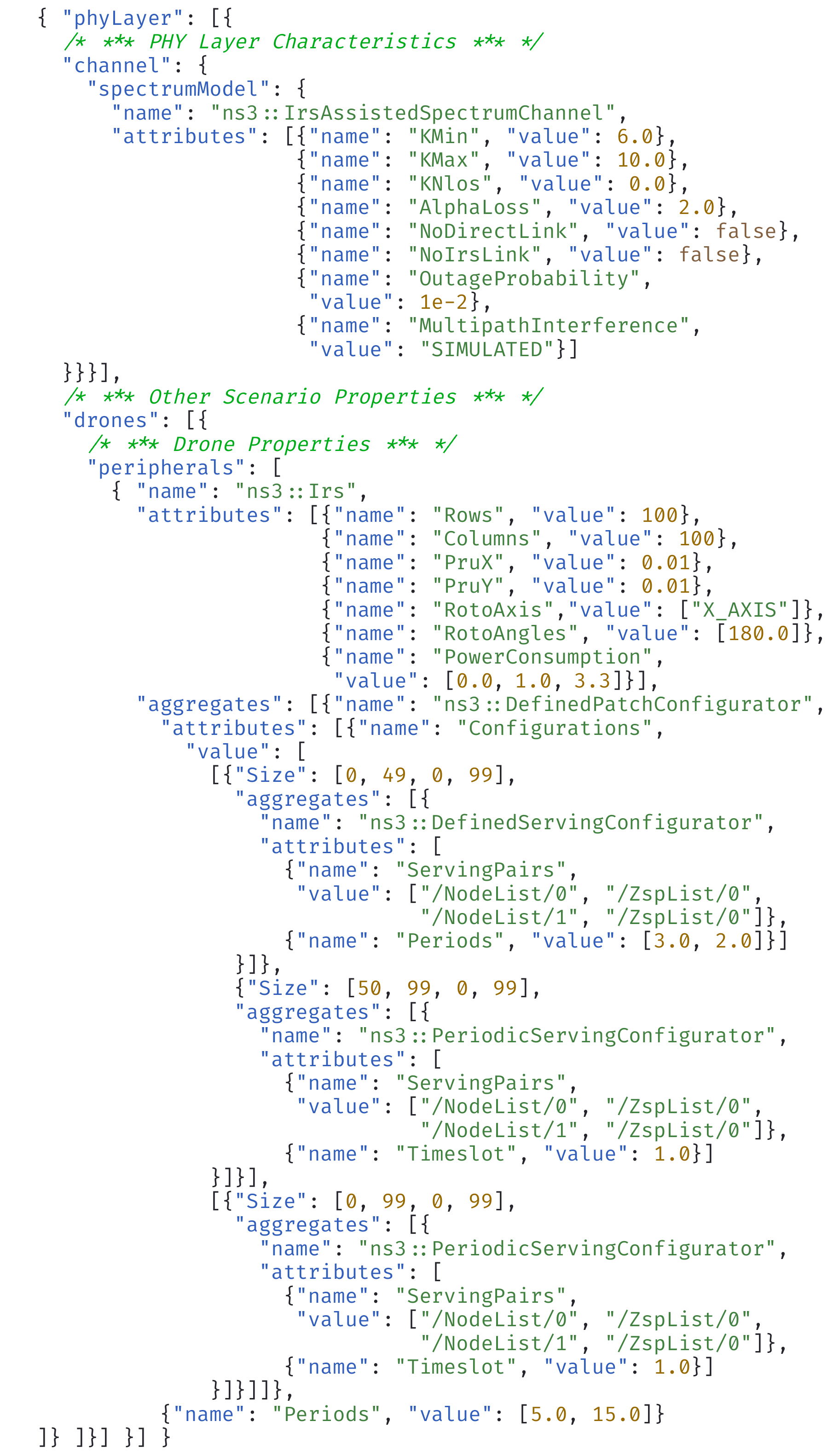}
    \caption{Extract of a \ac{JSON} scenario configuration that employs the \ac{IRS}-aware spectrum channel model and installs a single \ac{IRS} on a drone to serve a pair of nodes.}
    \label{fig:jsonconf}
\end{figure}

In order to provide a flexible and dynamic configurations at runtime, the proposed implementation offers additional configurator classes \texttt{ns3::PatchConfigurator} and \texttt{ns3::ServingConfigurator}. The former sets up the number and size of \acp{IRS} patches, called \textit{Patch Configurators}. The latter schedules the nodes to be served by each patch, namely \textit{Serving Configurators}.

For what concerns \textit{Patch Configurators}, the \texttt{ns3::DefinedPatchConfigurator} represents a basic reference already available in the module.
\begin{table}[!ht]
\centering
\scriptsize
\begin{tabular}{|p{2.45cm}|p{1.7cm}|p{1.6cm}|p{1.45cm}|}
\hline
\textbf{Parameter} & \textbf{Value} & \textbf{Parameter} & \textbf{Value} \\ \hline
Simulated Area & $400 \times \SI{400}{[m^2]}$ & \texttt{KMax} & $\SI{10}{[dB]}$ \\ \hline
\texttt{KMin} & $\SI{6}{[dB]}$ & \texttt{KNlos} & $\SI{0}{[dB]}$ \\ \hline
\texttt{AlphaLoss} & $\{3,\,4\}\ \si{[\#]}$ & \texttt{NoIrsLink} & \texttt{false} \\ \hline
\texttt{OutageProbability} & $\SI{0.01}{[\#]}$ & \texttt{NoDirectLink} & \texttt{false} \\ \hline
\texttt{RotoAxis} & \texttt{["X\_AXIS"]} & \texttt{RotoAngles} & $[180.0]\ \si{[deg]}$ \\ \hline
\texttt{PruX}, \texttt{PruY} & $\SI{0.01}{m}$ & \acs{UE}, \acs{eNB} Power & $24,\,\SI{49}{[dBm]}$ \\ \hline
\end{tabular}
\caption{Parameter settings.}
\label{tab:simparams}
\end{table}
It allows the definition of different patch setups that the \ac{IRS} adopts over time, as depicted in Figure \ref{fig:defined_patch_configurator}. It can be observed that the simulation starts by dividing the \ac{IRS} in two parts, as specified by \textit{Configuration 0}. At time $t_0$, the patches are reorganized to follow the map given by \textit{Configuration 1}. This logic reiterates twice more, until the end of the simulation.

\begin{figure*}[!ht]
    \centering
    \includegraphics[width=0.99\textwidth]{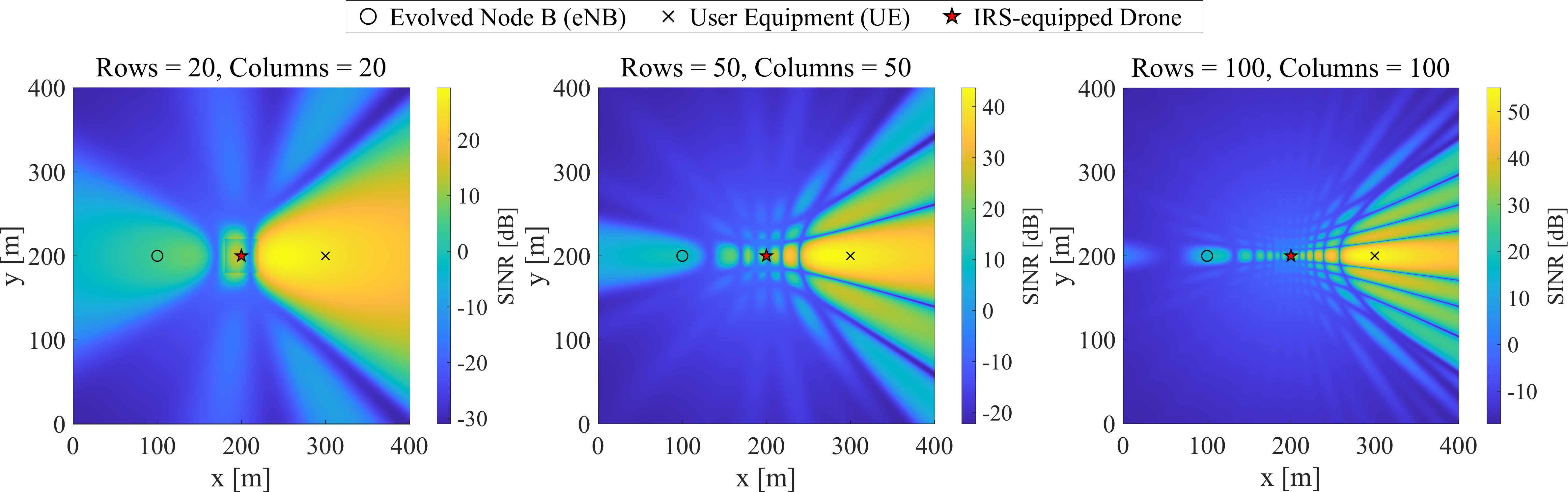}
    \caption{Comparison among downlink \acp{REM} for different \ac{IRS} sizes.}
    \label{fig:s1_dim_var}
\end{figure*}

Regarding the nodes to be served, instead, they can be scheduled according to one of the available \textit{Serving Configurator} algorithms: \texttt{ns3::DefinedServingConfigurator} which enables the definition of a list of node pairs to assist for different time intervals; \texttt{ns3::PeriodicServingConfigurator} which schedules node pairs in round-robin fashion for the same amount of time; \texttt{ns3::RandomServingConfigurator} randomly chooses which node pair to assist for a fixed time interval.

The \textit{Simulation Development Platform} allows an easy configuration of (i) the channel model properties, (ii) the \acp{IRS} setup, and (iii) the scheduling plans through a \ac{JSON} file, as shown in Figure \ref{fig:jsonconf}.
Such feature is enabled by the \texttt{ns3::ModelConfigurationVector} and \texttt{ns3::ModelConfigurationMatrix}, which have been developed as an extension in the \textit{Scenario Configuration Interface}, in order to dynamically apply different configurations at runtime.
As it can be noticed, the channel model can be configured through the parameters that are described in Section \ref{sec:chmod}, where \texttt{OutageProbability} is $\epsilon$, \texttt{KMin} is $\kappa^{\text{\tiny MIN}}$, \texttt{KMax} is $\kappa^{\text{\tiny MAX}}$, and \texttt{AlphaLoss} is $\alpha$. Furthermore, \texttt{NoDirectLink} and \texttt{NoIrsLink} represent booleans useful to analyze use cases where the direct and the reflected links are suppressed.
\texttt{MultipathInterference}, instead, can assume three different values which affect the interference introduced by the direct and reflected links (the second cosine in Equation \eqref{eq:nug}): \texttt{DESTRUCTIVE} for a purely destructive interference (i.e., worst case), \texttt{SIMULATED} for the actual one, and \texttt{CONSTRUCTIVE} for no interference at all (i.e., best case).
Finally, \texttt{KNlos} is the K-factor adopted when the direct link is in \ac{NLoS}.
Such parameters can be further tuned to simulate better or worse channel conditions according to the simulation design requirements.

\ac{IRS} configuration can be declared in the \ac{JSON} as a drone peripheral. In the example given in Figure \ref{fig:jsonconf}, two configurations are applied to the \ac{IRS}, with different time durations, specified in \textit{Periods}. In the first one, the \ac{IRS} is split in half with two patches: one patch assists the links of two \ac{GU}, with global index 0 and 1, for three and two seconds, respectively; the second one periodically serves the same users for one second each. Further, in the second configuration, the whole \ac{IRS} is used for 15 seconds to serve both nodes in round-robin fashion for one second each. Finally, a power consumption, related to the \ac{IRS} controller, is also defined.

\section{Simulation Campaign}\label{sec:simcampaign}
The entire simulation campaign revolves around the features studied for both the mathematical model, described in Section \ref{sec:chmod}, and the \ac{IoD-Sim} implementation, presented in Section \ref{sec:swdes}. Three different scenarios are designed and assessed hereby to validate the features introduced by the \ac{IRS} module, adopting the parameters reported in Table \ref{tab:simparams}, if not otherwise specified. Furthermore, all the scenarios are tested using one communication technology only, i.e., \ac{LTE}, with a fixed bandwidth of $\SI{5}{MHz}$ and 25 resource blocks. In particular, all these scenarios consider a \ac{eNB}, acting as a \ac{ZSP}/\ac{BS}, and a set of \acp{UE}, acting as nodes/\acp{GU}, that experience different \ac{SINR} levels due to pathloss and \ac{LoS} conditions.
To this end, \ac{IRS}-equipped \acp{UAV} are employed to assist the communication links. The overall performance achieved through the aid of the \ac{IRS} are compared, analyzed, and discussed via several \acp{KPI}, such as \ac{REM}, \ac{SINR}, maximum achievable rate, and average throughput.

\subsection{Scenario \#1}
The first Scenario considers an area with a building of $20\times 20\times\SI{25}{m^3}$, placed at $[200,200,0]$, that obstructs the direct link between an \ac{eNB} and a \ac{UE}, located at $[100,200,30]$ and $[300,200,0]$, respectively. To support the communication between these two nodes, an \ac{IRS}-equipped \ac{UAV} hovers $\SI{50}{m}$ over the building, thus re-establishing the \ac{LoS}.
The overall Scenario is depicted in Figure \ref{fig:s1_dim_var}, which also illustrates the downlink \ac{REM} at the ground level with a resolution of $\SI{16}{samples/m^2}$. It is worth specifying that, for the sake of the analysis, the contribution of the \ac{eNB}-\ac{UE} direct link is temporarily neglected. The radiation fingerprints exhibit two main properties: as the number of \acp{PRU} increases (i) the main lobe pointing at the target node becomes narrower and (ii) the perceived \ac{SINR} increases as well. However, these benefits come at the price of a larger \ac{IRS}, which implies higher costs and footprint.

\begin{figure}[!t]
    \centering
    \includegraphics[width=\columnwidth]{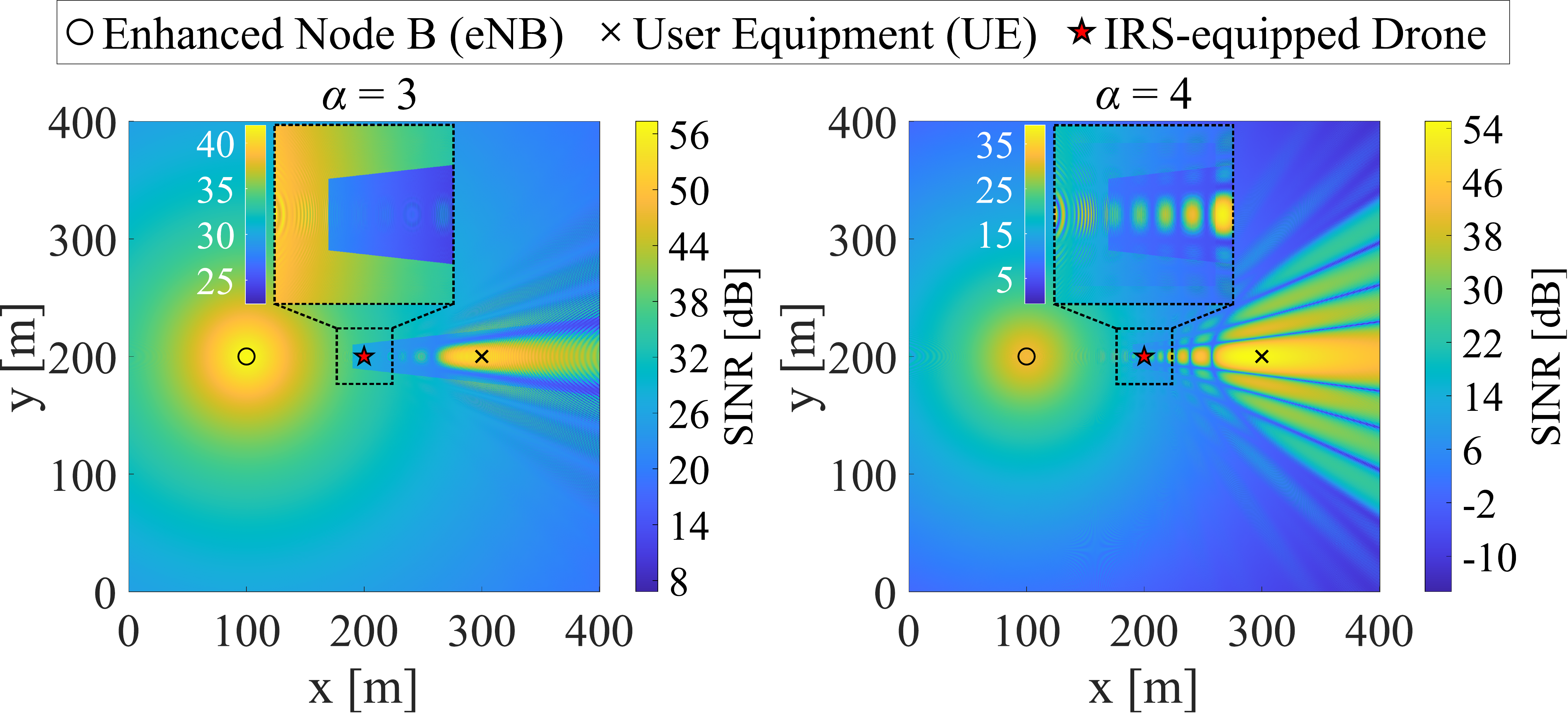}
    \caption{Downlink \acp{REM} for different attenuation factors.}
    \label{fig:s1_3vs4}
\end{figure}

Figure \ref{fig:s1_3vs4} depicts the same scenario from a different point of view. The direct link is not neglected anymore and it is considered an \ac{IRS} of fixed size $100 \times 100$ elements, with a varying attenuation factor $\alpha = \{3,\,4\}$ adopted for the direct link.
Clearly, this case highlights the shadowing effect due to the presence of the building, which is more evident with $\alpha = 3$, since the direct link is less attenuated. At the same time, with $\alpha = 4$, it is more noticeable a weaker shadow surrounding the building, which is its projection on the ground as a result of the \ac{IRS} reflections, i.e., \ac{UAV}-\ac{UE} \ac{NLoS} link. Moreover, it can be noticed a slight ripple effect due to fast-fading phenomena as a consequence of multipath interference.

\begin{figure}[!t]
    \centering
    \includegraphics[width=\columnwidth]{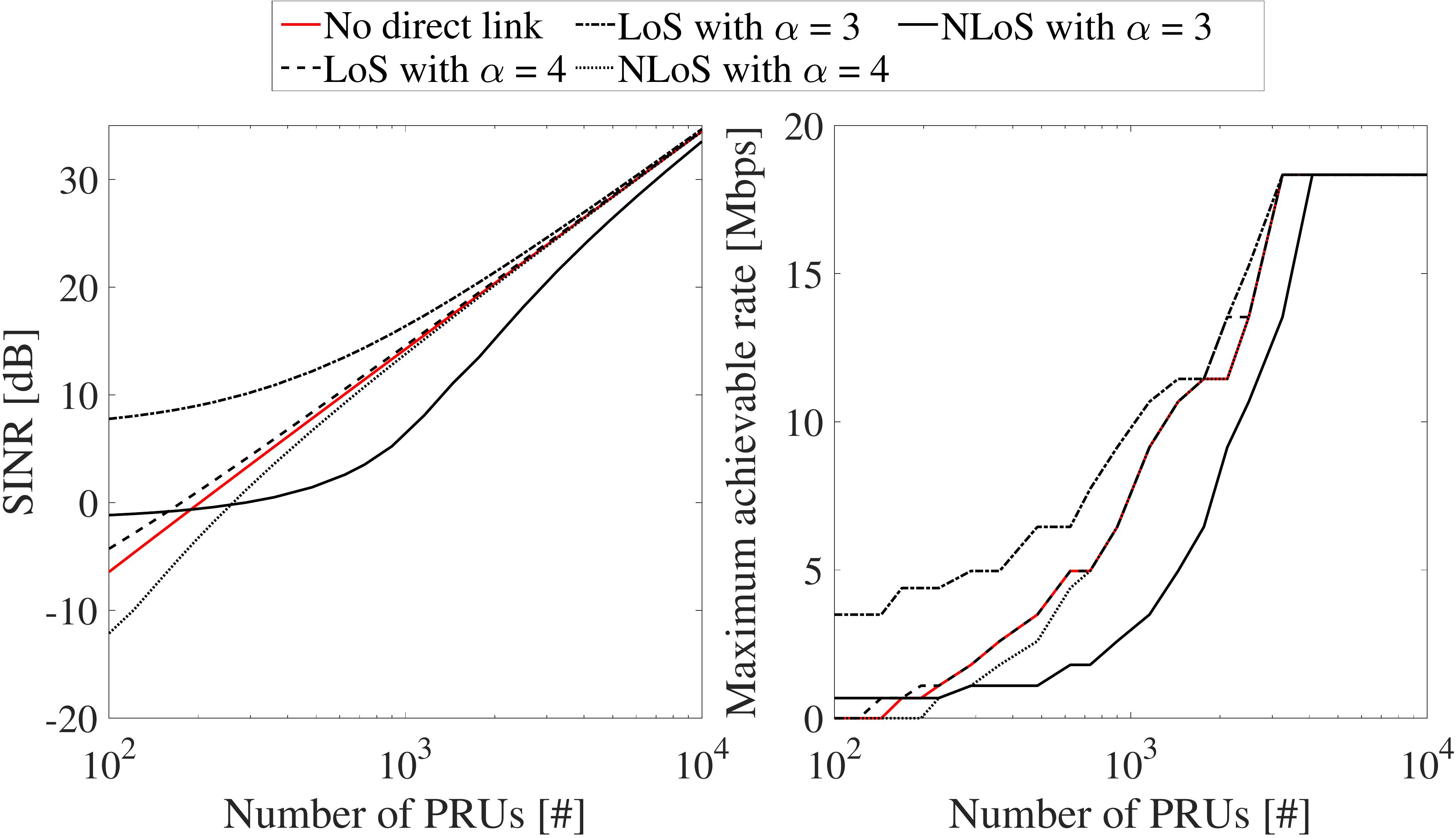}
    \caption{Uplink maximum achievable rate and \ac{SINR} under different channel conditions.}
    \label{fig:s1_rate_sinr}
\end{figure}

Lastly, Figure \ref{fig:s1_rate_sinr} shows the channel conditions between the \ac{eNB} and \ac{UE} in terms of \ac{SINR} and maximum achievable data rate in uplink. This time, multiple configurations investigate the presence and also the absence of the building, labeled with ``\acs{LoS}'' and ``\acs{NLoS}''. Moreover, the total absence of the \ac{eNB}-\ac{UE} direct link is considered, marked as ``No direct link''. In terms of \ac{SINR}, illustrated on the left, the first obvious observation is that, for a given $\alpha$, the \ac{LoS} cases are always better than the \ac{NLoS} ones.
Moreover, for a low number of elements, the curves with $\alpha = 3$ start with a better \ac{SINR} with respect to the ones characterized by $\alpha = 4$. However, as the \ac{IRS} becomes larger, the less attenuated case in \ac{NLoS} conditions, i.e., $\alpha = 3$, is characterized by a significant destructive interference phenomena, as it can be noticed by comparing it with the ''No direct link`` curve. These unwanted effects can be prevented with a proper design of the scenario geometry, i.e., the communication actors should be correctly aligned. Lastly, for a very large number of elements, all cases converge to ''No direct link``, since the reflected link overshadows the direct one. For what concerns the maximum achievable rate, depicted on the right, it follows a trend that is similar to the \ac{SINR}. It can be observed that, as the number of \acp{PRU} grows, the curves overlap due to \ac{MCS} switching \cite{3gpp.36.213}, until the rate saturates at $\SI{18.336}{Mbps}$.

\begin{figure*}[!t]
    \includegraphics[width=0.99\textwidth]{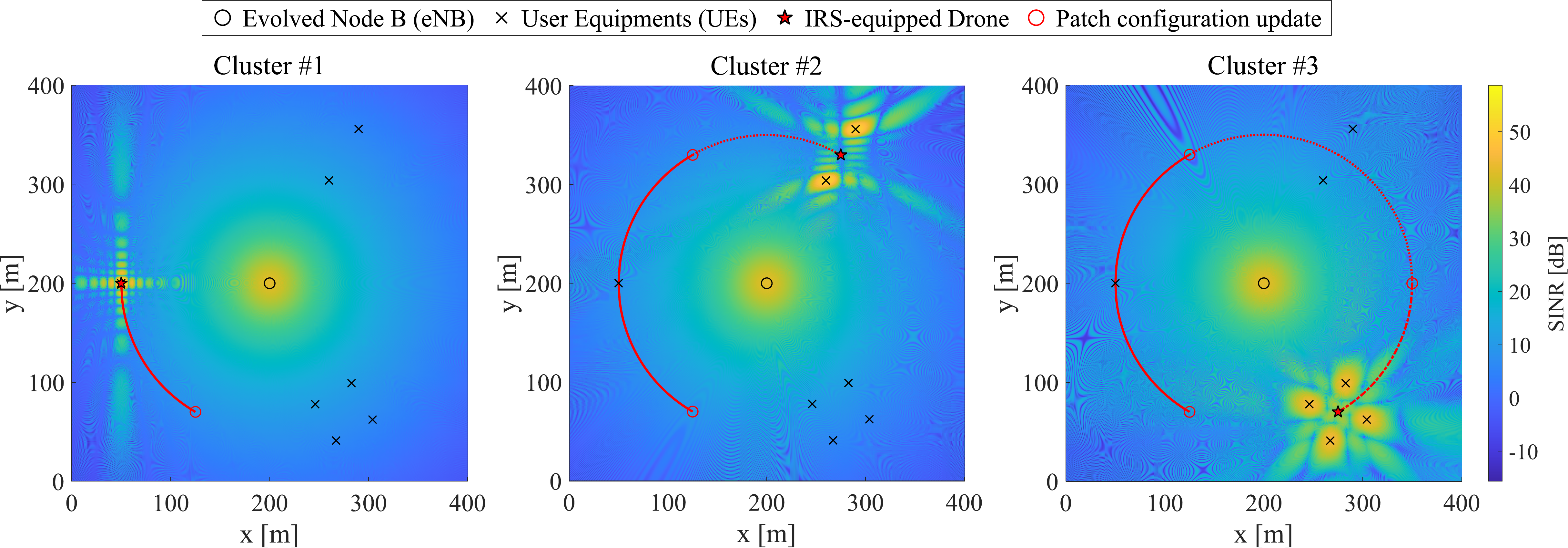}
    \caption{\acp{REM} taken exactly when the drone results orthogonal to each cluster.}
    \label{fig:s2_mission}
\end{figure*}
\begin{figure*}[!t]
    \centering
    \includegraphics[width=0.99\textwidth]{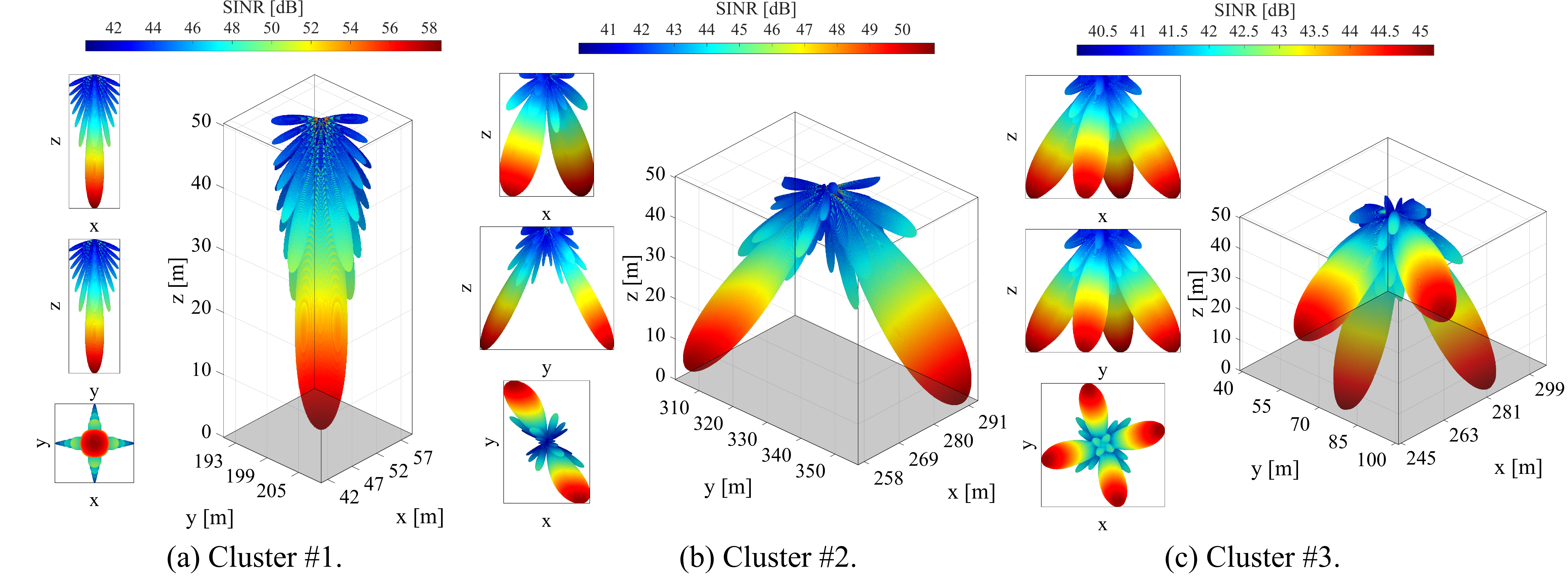}
    \caption{The 3D radiation patterns of the \ac{IRS} serving different clusters.}
    \label{fig:s2_beams}
\end{figure*}
\begin{figure}[t]
    \centering
    \includegraphics[width=0.49\textwidth]{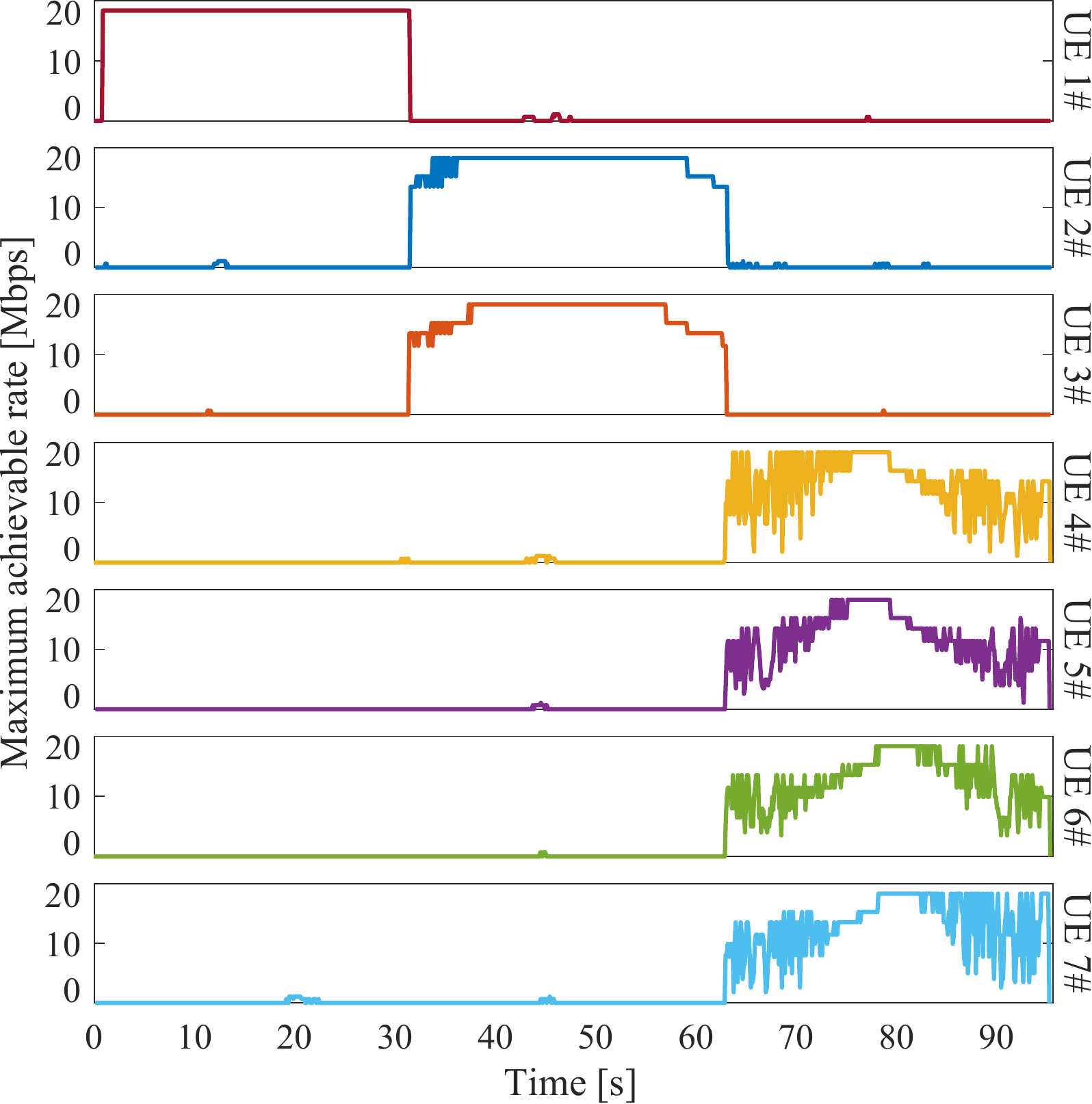}
    \caption{Uplink maximum achievable rate.}
    \label{fig:s2_rate}
\end{figure}

\subsection{Scenario \#2}
This scenario considers the presence of three clusters: (i) Cluster \#1 is made by 1 \ac{UE} placed in $[50, 200, 0]$, (ii) Cluster \#2 has 2 \acp{UE} positioned at $\{[260, 303.923, 0],[290, 355.885, 0]\}$, and (iii) Cluster \#3 is characterized by 4 \acp{UE} located at $\{[282.765, 99.074, 0]$, $[303.978, 62.331, 0]$, $[267.235, 41.118, 0]$, $[246.022, 77.86, 0]\}$.
All the \acp{UE} exchange data with an \ac{eNB} with the support of an \ac{IRS}-equipped \ac{UAV}. The direct \ac{UE}-\ac{eNB} link is characterized by the pathloss exponent $\alpha=4$. The goal is to fairly serve each cluster through a \ac{IRS} of $100 \times 100$ elements. To this end, the drone follows a circular trajectory of radius $\SI{150}{m}$, at a constant speed of $\SI{10}{m/s}$, that intersects the center of each cluster. The circumference is equally divided into three arcs, for which a suitable \ac{IRS} configuration is set to serve the \acp{UE} of interest for $\sim \SI{31.416}{s}$.

The described scenario is depicted in Figure \ref{fig:s2_mission}, which also shows the downlink \ac{REM} at three different instants corresponding to the \ac{UAV} being orthogonal to the center of each cluster.
As it can be seen, the reflected signal power yields a different radiation pattern on the ground, depending on the adopted \ac{IRS} configuration. As already seen in Scenario \#1, this case is subject to the interference between direct and reflected links, i.e., multipath. Additionally, this effect is exacerbated by the presence of patches configured to serve different members of the same cluster, i.e., the \ac{IRS} self-interference.
Furthermore, the \ac{SINR} is inversely proportional to the number of users to be served. This behavior is more evident in Figure \ref{fig:s2_beams}, where the signal beams produced by the \ac{IRS} are depicted, with a peak \ac{SINR} of $\sim \SI{58.61}{dB}$ for Cluster \#1, $\sim \SI{50.98}{dB}$ for Cluster \#2, and $\sim \SI{45.19}{dB}$ for Cluster \#3. Specifically, the \ac{SINR} lowers since the surface is equally divided among the nodes of the target cluster. It is worth noting that, differently from the Clusters \#1 and \#3, the two beams depicted in Figure \ref{fig:s2_beams}b are not symmetrical, as can be seen in the x-z and y-z projections, due to the rectangular shape of the patches.

For the sake of completeness, in  Figure \ref{fig:s2_rate} it is investigated the uplink maximum achievable rate for all the \acp{UE}, since it is more critical with respect to the downlink one.
The contribution of the \ac{UAV} is crucial to allow the communication between these nodes and the \ac{eNB}. Indeed, when the \acp{UE} are no longer supported by the drone, the data rate drops to zero due to the high loss characterizing the direct link. Otherwise, it can be observed time-discrete variations of the rate, which are caused by the \ac{MCS} switching. This is due to (i) the variation of the \ac{UAV}-\acp{GU} distance and (ii) the fast-fading effect, which clearly worsens as the number of served \acp{UE} increases.
Moreover, when the \ac{UAV} is closest to a \ac{UE}, the theoretical maximum rate of $\SI{18.336}{Mbps}$ is reached, as already seen in Figure \ref{fig:s1_rate_sinr} of Scenario \#1.

\subsection{Scenario \#3}
\begin{figure}[t]
    \centering
    \includegraphics[width=0.49\textwidth]{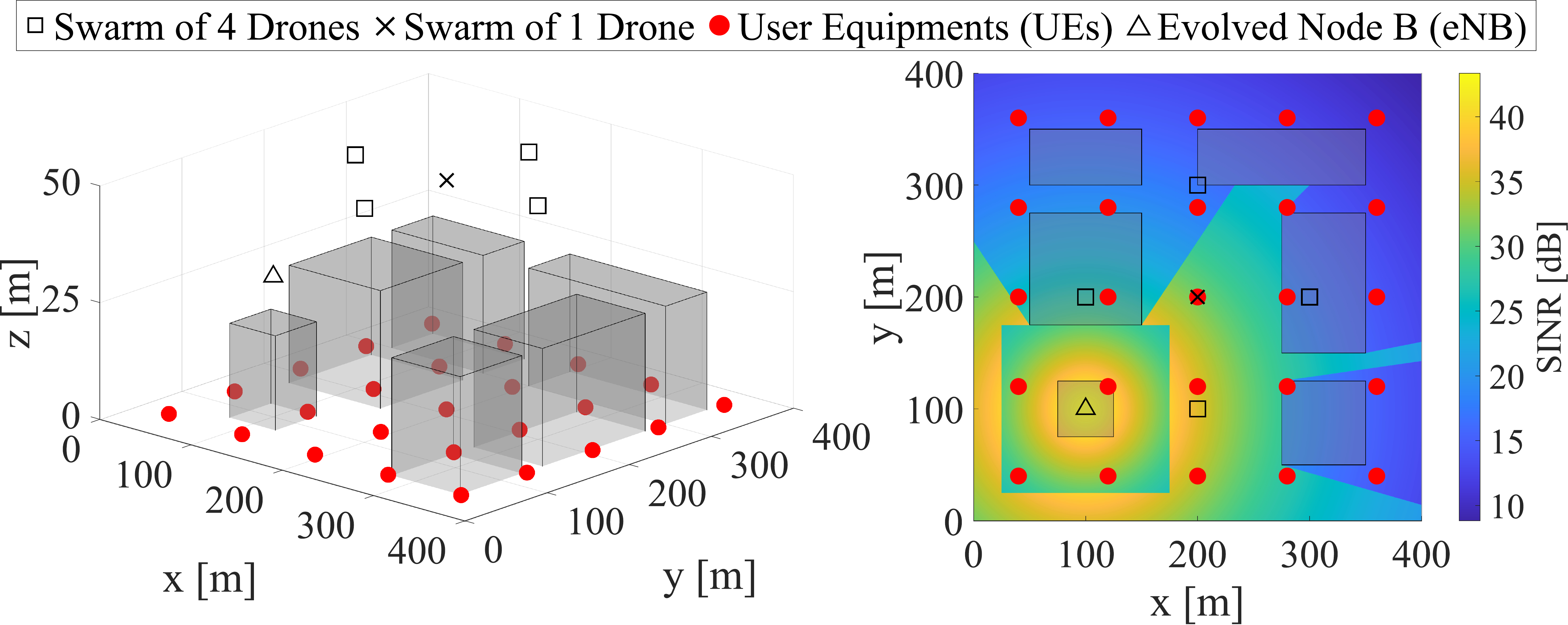}
    \caption{Simulated smart city scenario.}
    \label{fig:s3_design}
\end{figure}

The last scenario investigates all the available \textit{Serving Configurators} described in Section \ref{sec:swdes}, in the context of a smart city. Indeed, the urban environment is particularly useful to analyze both \ac{LoS} and \ac{NLoS} cases.
As depicted in the left of Figure \ref{fig:s3_design}, multiple buildings and an uniform grid of 25 \acp{UE} are considered.
Each \ac{UE} communicates with an \ac{eNB} placed on the top of the bottom-left building, at $\SI{30}{m}$ of height.
As it can be noted, in the right of Figure \ref{fig:s3_design}, the downlink \ac{REM} shows the shadowing effect due to the presence of buildings, which obstruct the \ac{LoS} between some \acp{UE} and the \ac{eNB}. As a consequence, there are nodes that cannot communicate, since the \ac{SINR} is under the threshold, according to the \texttt{ns3::MiErrorModel} \cite{mezzavilla2012lightweight}. To cope with this issue, the communication system is enhanced with one and then four \ac{IRS}-equipped \acp{UAV}. In the former case, the \ac{UAV} is placed in $[200, 200, 50]$, whereas in the latter the \acp{UAV} are located at $\{[100, 200, 50]$, $[200, 300, 50]$, $[300, 200, 50]$, $[200, 100, 50]\}$.

\begin{figure}[t]
    \centering
    \includegraphics[width=0.49\textwidth]{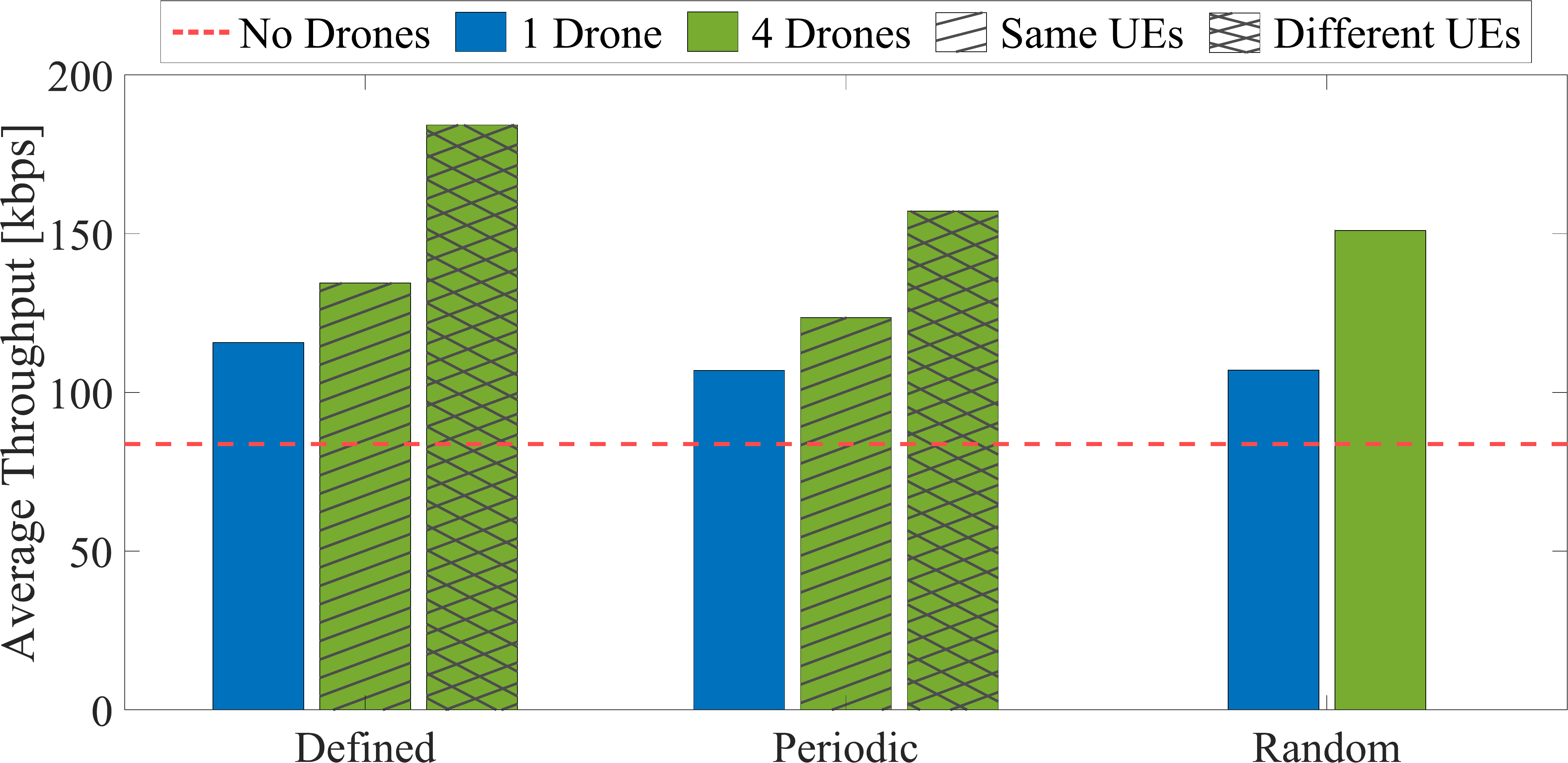}
    \caption{UEs average throughput adopting different \textit{Serving Configurators}.}
    \label{fig:s3_rate_small}
\end{figure}
\begin{figure}[t]
    \centering
    \includegraphics[width=0.49\textwidth]{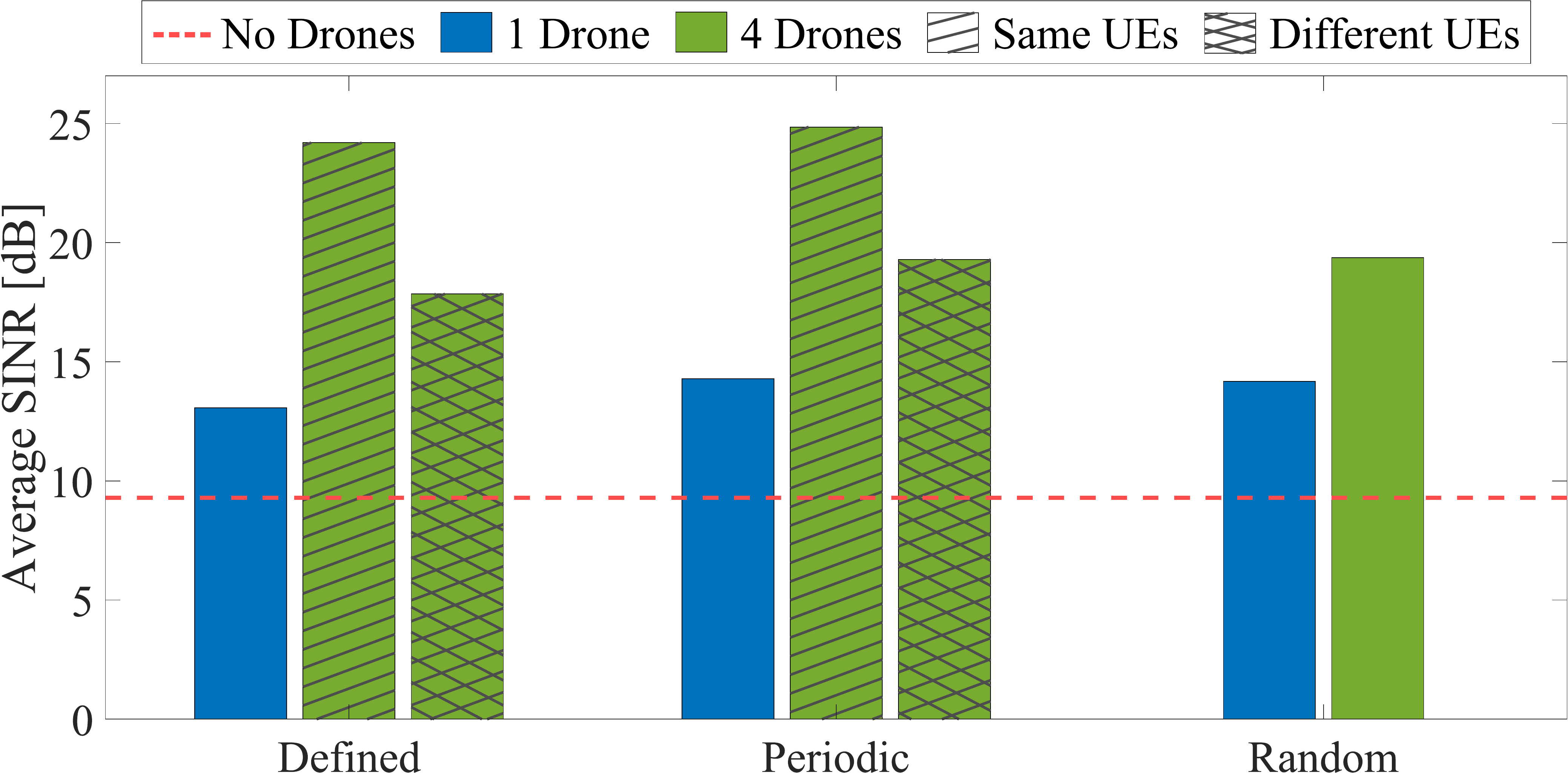}
    \caption{UEs average \ac{SINR} adopting different \textit{Serving Configurators}.}
    \label{fig:s3_sinr_small}
\end{figure}

In order to saturate the \ac{LTE} capacity, a live streaming traffic is simulated for a mission that lasts $\SI{75}{s}$. With this aim, the \texttt{ns3::UdpEchoClientApplication} is employed, which leverages the \ac{UDP} protocol to transmit a packet of $\SI{64}{KiB}$, i.e., the maximum possible size, every $\SI{0.03}{s}$.
All the three proposed \textit{Serving Configurators} described in Section \ref{sec:swdes} are tested, labeled as \textit{Defined}, \textit{Periodic}, and \textit{Random}. Specifically, the former is set up to exclusively serve the nodes that have an \ac{SINR} below the threshold. The last two, instead, serve all the nodes.
Moreover, a baseline is also considered (the red dashed line) in which no drones assist the \acp{UE}, which have to rely solely on the direct link with the \ac{eNB}.
The results, in terms of average throughput and \ac{SINR}, are reported in Figures \ref{fig:s3_rate_small} and \ref{fig:s3_sinr_small}, respectively.
In both Figures, the blue bars indicate the case where only one \ac{IRS}-equipped drone is employed, whereas the green ones consider a swarm of 4 \acp{UAV}. In the latter case, patterns are used to distinguish two different approaches: ``Same \acp{UE}'' refers to the case in which all drones simultaneously assist a given \ac{UE} in each time interval; vice versa, ``Different \acp{UE}'' indicates that all drones serve distinct \acp{UE}. For obvious reasons, the ``Random'' case do not discuss such a difference.

It is evident that, with respect to the baseline approach, the employment of \acp{IRS} leads to an improvement in both the average throughput and \ac{SINR}, of at least $\sim 27.66\%$ and $\sim 40.6\%$, respectively.
Of course, these benefits become more prominent as the number of drones increases.
Among the adopted configurators, it can be noted that there are no major differences in terms of \ac{SINR}. Indeed, even if in different orders, \acp{UE} are served for about the same time and with the same bandwidth. Nonetheless, the \textit{Periodic} presents slightly better performances. However, when the average throughput is considered, the \textit{Periodic} configurator (which performs very similar to the random one) does not guarantee the same benefits brought by the \textit{Defined} one.
Indeed, the latter focuses on serving those nodes which demand more signal power to reach the required minimum \ac{SINR}, which in turn produces an higher overall system throughput.
A similar rationale can be applied when, given a configurator, ``Same \acp{UE}'' and ``Different \acp{UE}'' are compared. In fact, serving distinct \acp{UE} at the same time allows them to use a higher \ac{MCS}, which yields a greater average throughput, even if the corresponding \ac{SINR} are comparable.

\section{Conclusions}\label{sec:conclusions}
The \ac{IoD} represents a major leap in telecommunications, as it enables on-demand network coverage and a high degree of versatility. At the same time, \acp{IRS} allow to control the environmental conditions of the radio channel, thus leading to noticeable improvements in communication quality. As drones and \acp{IRS} clearly represent key-enablers for 6G communications, this work proposes a module based on the \ac{IoD-Sim} platform, which enables the development of future communication systems where this two technologies can be integrated. This module represents a flexible solution thanks to the available general schedulers for \ac{IRS} patches.

Despite the manifold features already available, the \ac{IRS} module will be improved in the future, with more efforts focused on:
\begin{enumerate}
    \item Accurate power consumption model of \acp{IRS}.
    \item Performance assessment of this module with mmWave simulations, to assess the performance of systems that go beyond classical sub-6GHz communications.
    \item Comparison of these new emerging systems with \ac{AF} solutions, employed in 5G network backhaul.
    \item Real-time attitude controls for the \ac{IRS} in order to follow a target mobile node.
    \item Enhanced configurators with feedback loop that choose to serve nodes depending on their channel conditions.
    \item Channel model aware of the specific material obstructing the \ac{LoS}, in order to choose the most suitable K-factor and pathloss coefficient.
\end{enumerate}
Hopefully, this work will stimulate the scientific community to improve the intelligence of these emerging devices in multiple directions. The emergence of a flourishing and empowering community on open-source collaboration platforms will ultimately determine the success of future development endeavors.

\bibliographystyle{IEEEtran}
\bibliography{bibliography}
\begin{IEEEbiography}
    [{\includegraphics[width=1in,height=1.25in,clip,keepaspectratio]{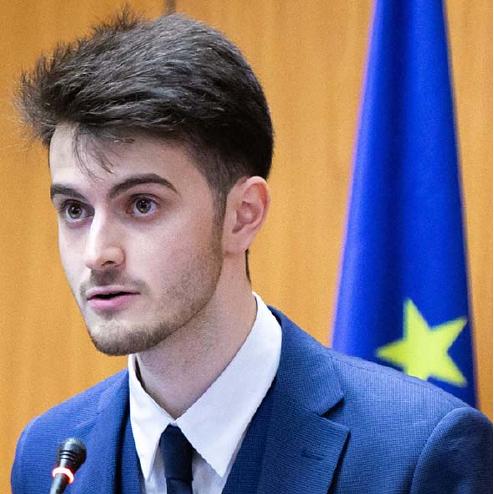}}]{Giovanni Grieco}
received the Dr. Eng. degree (with honors) in Telecommunications Engineering from Politecnico di Bari, Bari, Italy in October 2021. His research interests include Internet of Drones, Cybersecurity, and Future Networking Architectures. He is the principal maintainer of IoD\_Sim. Since 2021, he has been a Ph.D. student at the Department of Electrical and Information Engineering at Politecnico di Bari.
\end{IEEEbiography}
\begin{IEEEbiography}[{\includegraphics[width=1in,height=1.25in,clip,keepaspectratio]{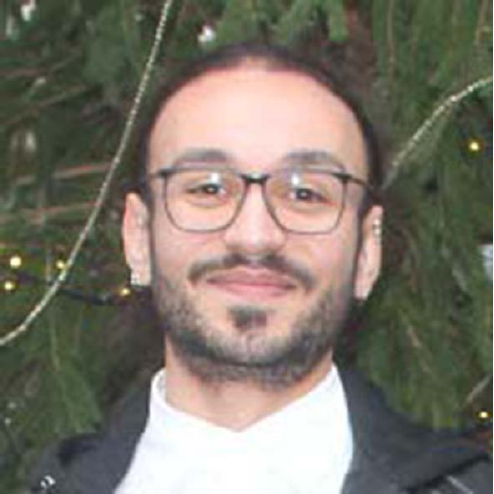}}]{Giovanni Iacovelli}
received the Ph.D. in electrical and information engineering from Politecnico di Bari, Bari, Italy, in December 2022. His research interests include Internet of Drones, Machine Learning, Optimization and Telecommunications. Currently, he is a Research Fellow at the Department of Electrical and Information Engineering, Politecnico di Bari.
\end{IEEEbiography}
\begin{IEEEbiography}
    [{\includegraphics[width=1in,height=1.25in,clip,keepaspectratio]{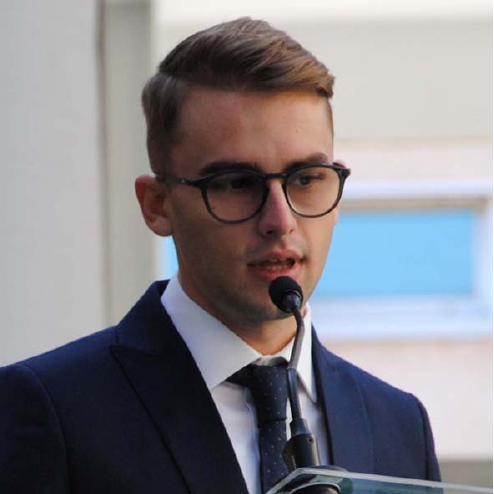}}]{Daniele Pugliese}
received his Bachelor's degree (with honors) in Computer and Automation Engineering in December 2021 from Politecnico di Bari, Bari, Italy, where he is currently a Master's Student in Telecommunications Engineering. Since March 2022 he is a research fellow at the Department of Electrical and Information
Engineering, Politecnico di Bari.
\end{IEEEbiography}
\begin{IEEEbiography}
    [{\includegraphics[width=1in,height=1.25in,clip,keepaspectratio]{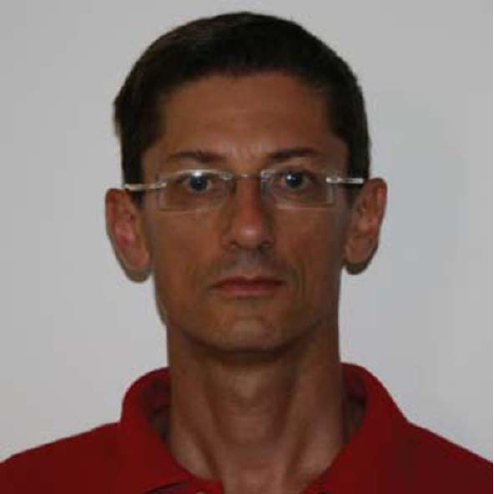}}]{Domenico Striccoli}
received, with honors, the Dr. Eng. Degree in Electronic Engineering in April 2000, and the Ph.D. degree in April 2004, both from the Politecnico di Bari, Italy. In 2005 he joined the DIASS Department of Politecnico di Bari in Taranto as Assistant Professor in Telecommunications. Actually he teaches fundamental courses in the field of Telecommunications at Department of Electrical and Information Engineering (DEI) of Politecnico di Bari, where he holds the position of Assistant Professor in Telecommunications.
Dr. Striccoli has authored numerous scientific papers on various topics related to electronic engineering. His work has been published in esteemed international journals and presented at leading scientific conferences.
\end{IEEEbiography}
\begin{IEEEbiography}[{\includegraphics[width=1in,height=1.25in,clip,keepaspectratio]{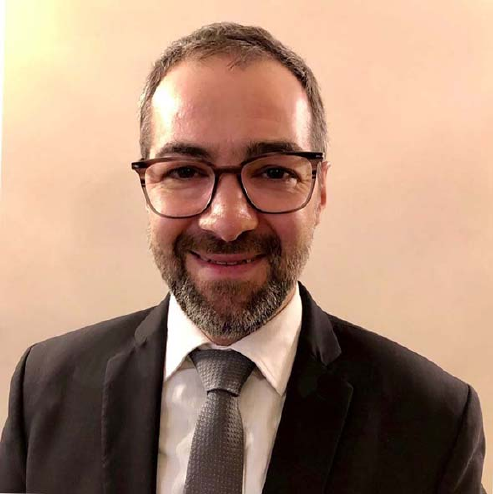}}]{L. Alfredo Grieco}
is a full professor in telecommunications at Politecnico di Bari. His research interests include Internet of Things, Future Internet Architectures, and Nano-communications. He serves as Founder Editor in Chief of the Internet Technology Letters journal (Wiley) and as Associate Editor of the IEEE Transactions on Vehicular Technology journal (for which he has been awarded as top editor in 2012, 2017, and 2020).
\end{IEEEbiography}
\end{document}